\documentclass[dvips,preprint,aps,prb,nofootinbib,endfloats*]{revtex4}
\usepackage{graphicx}
\usepackage{amsmath,amssymb,amsfonts,mathrsfs}
\begin{document}
\frenchspacing
\newcommand{\Wc}{W$_\textrm{1-3}$}
\newcommand{\Wcc}{W$_\textrm{1-5}$}
\newcommand{\chair}{$^4$C$_1$}
\newcommand{\preprintclearpage}{\clearpage}
\newcommand{\sso}{\hspace{\stretch{1}}}
\newcommand{\invchair}{$^1$C$_4$}
\newcommand{\boatH}{$^{3,\mathrm{O}}$B}
\newcommand{\boatL}{B$_{3,\mathrm{O}}$}
\renewcommand{\skew}{$^3$S$_1$}

\author{E.~Autieri}
\author{M.~Sega}
\email{sega@science.unitn.it}
\author{F.~Pederiva}
\affiliation{Department of Physics and I.N.F.N., University of Trento, via Sommarive 14, 38123 Trento, Italy}
\author{G.~Guella}
\affiliation{Department of Physics, University of Trento, via Sommarive 14, 38123 Trento, Italy}

\title{Puckering Free Energy of Pyranoses: an NMR and
Metadynamics--Umbrella Sampling Investigation}

\begin{abstract}
We present the results of a combined metadynamics--umbrella sampling investigation of
the puckered conformers of pyranoses described using the {\sc{}gromos} 45a4 force field.
The free energy landscape of Cremer--Pople puckering coordinates has been calculated for
the whole series of $\alpha$ and $\beta$ aldohexoses, showing that the
current force field parameters fail in reproducing proper puckering free energy differences between chair conformers. 
We suggest a modification to the {\sc{}gromos} 45a4 parameter set which  improves
considerably the agreement of simulation results with theoretical and experimental
estimates of puckering free energies. We also report on the experimental measurement of
altrose conformers populations by means of NMR spectroscopy, which show good agreement 
with the predictions of current theoretical models.

\end{abstract}
\maketitle
\section{Introduction}
Within the framework of classical force fields, the number of computer experiments on
saccharides has grown considerably in recent years, and various systems have been
addressed\cite{hardy93a,hardy96a,almond98a,ueda98a,venkatarangan99a,lee00a,cavalieri01a,naidoo01a,kim01a,momany02a,paradossi02a,paradossi02b,lawtrakul03a,umemura04a,verli04a,yu04a,sandoval05a,xie05a,gonzalez-outeirino05a,palleschi05a,Neelov06,mamonova06a,figueiras07a,kony07a,yoshida08a,almlof08a,umemura09a}. Devising a realistic model of monosaccharides is obviously a decisive step in order for
carbohydrates simulations to have enough predictive power. The accurate description of
monosaccharides with classical force fields is not an easy task, because of the
delicate interplay of different factors such as the presence of a high number of intramolecular
hydrogen bonds, the competition of these hydrogen bonds with water-sugar ones and important
steric and electrostatic effects between ring substituents in spatial proximity (see for
example Ref.~\onlinecite{Krautler07} and references within). The problem of reproducing some
carbohydrates peculiarities, such as the rotameric distribution of the hydroxymethyl group
or the anomeric and exo-anomeric effects, have been addressed in various force fields, and
the reader can find some comparative analyses in Refs.~\onlinecite{perez98a,behler01a,corzana04a}.
However, considerably less attention has been devoted so far to the correct reproduction of
the ring conformational properties.

Cyclic monosaccharides can appear indeed as puckered rings, and their conformational transitions
dramatically alter the equilibrium properties of both single sugar rings as well as those
in oligo and polysaccharides\cite{Kroon-Batenburg97}.  Despite the large number of
possible puckered conformers, many biologically relevant monosaccharides in the pyranoid
form appear almost always in one stable puckered conformation, the second most populated
state being so unlikely not to be detectable by actual experimental techniques. In spite of
that, several authors reported an inappropriately high percentage of secondary puckered
conformations\cite{brady86a,Kroon-Batenburg97,Spieser99,Lins05,chiessia,Limbach08a} when
modeling carbohydrates using classical force fields such as {\sc
gromos}\cite{Koehler87,Gunsteren96,Kouwijzer95,Ott68,Spieser99,Lins05} or {\sc
opls-aa}\cite{damm97a} for simulations of sugars in solution. 
Regarding the latest {\sc gromos} parameter set for carbohydrates (45a4)\cite{Lins05}, 
non-chair conformers have been shown to be accessible during equilibrium simulation runs 
of $\beta$\textsc{-d-}glucose\%\cite{Krautler07}, and two recent
works\cite{hansen10a,spiwok10a} estimated the free energy difference between chair 
conformers to be at least 10 kJ/mol lower than most theoretical and ab-initio simulation results.
Besides equilibrium simulations, the importance of ring puckering has been
proven by simulated pulling experiments, employed to interprete single molecule force-spectroscopy
data\cite{Lee04,Neelov06,heymann99a}. In this case, ring conformational transitions simulated
using three different force fields (AMBER94\cite{cornell95a},AMBER-GLYCAM\cite{woods95a} and
CHARMM-Parm22/SU01\cite{eklund03a}) led to different interpretation of the same experimental
data\cite{Neelov06}. Having control on the reliability of force fields in reproducing
puckering properties would certainly improve the predictivity of computer
simulations of saccharides.

It is hence understandable how desirable is a reparametrization of
those force fields which present unphysical ring conformers. This
difficult task, however, is made even more arduous by the lack of
experimental estimates of conformers populations, as the only
monosaccharide investigated so far is idose\cite{Snyder86}.
This work aims at filling, at least partially, these gaps. On one hand we
estimated the chair conformers populations of altrose, obtained from the
assignement of its $^1$H NMR spectrum. On the other hand, we performed 
a combined metadynamics -- umbrella sampling investigation\cite{Babin06} of the
puckering properties of all $\alpha$ and $\beta$\textsc{-d-}aldopyranoses, modeled
using the {\sc gromos} 45a4 parameter set\cite{Lins05}, and identified
a new set of parameters that satisfactorily reproduces the puckering free energies 
of the main ring conformers.

This paper is organized as follows: in Sec.~\ref{sec:problem} the problems
related to the determination of puckering free energy landscape will be
presented; in Sec.~\ref{sec:nmr} the results of an NMR investigation of altropyranose
conformers populations are presented; in
Sec.~\ref{sec:methods} the combined metadynamics--umbrella sampling will be discussed, as
well as the use of Cremer--Pople puckering coordinates\cite{Cremer75,Cremer90}, their 
peculiarities as collective variables, and the simulation details; 
in Sec.~\ref{sec:results} the puckering free energy profile
obtained with the {\sc gromos} 45a4 force field
for {$\beta$\textsc{-d-}glucose} is discussed in detail and compared to recent similar
calculations\cite{hansen10a,spiwok10a}.
In Sec~\ref{sec:parameters} we present the 
puckering free energy profiles of the whole $\alpha$ and $\beta$ series of \textsc{d-}aldohexoses
modeled with the {\sc gromos} 45a4 force field, showing that the puckering free energies
are so poorly reproduced, that the inverted chair conformers
of galactose, mannose and $\alpha$\textsc{-d-}Glc become significantly populated. Subsequently, we
propose a change to the 45a4 parameter set, showing how the free energies obtained
with the use of the new parameter set compare favorably with available theoretical and
experimental data. Eventually, some concluding remarks are presented in Sec. \ref{sec:conclusions}.

\section{The problem of pyranoses puckering free energy\label{sec:problem}}
\begin{figure}[t]
\centering
\smallskip
\includegraphics[width=0.9\columnwidth]{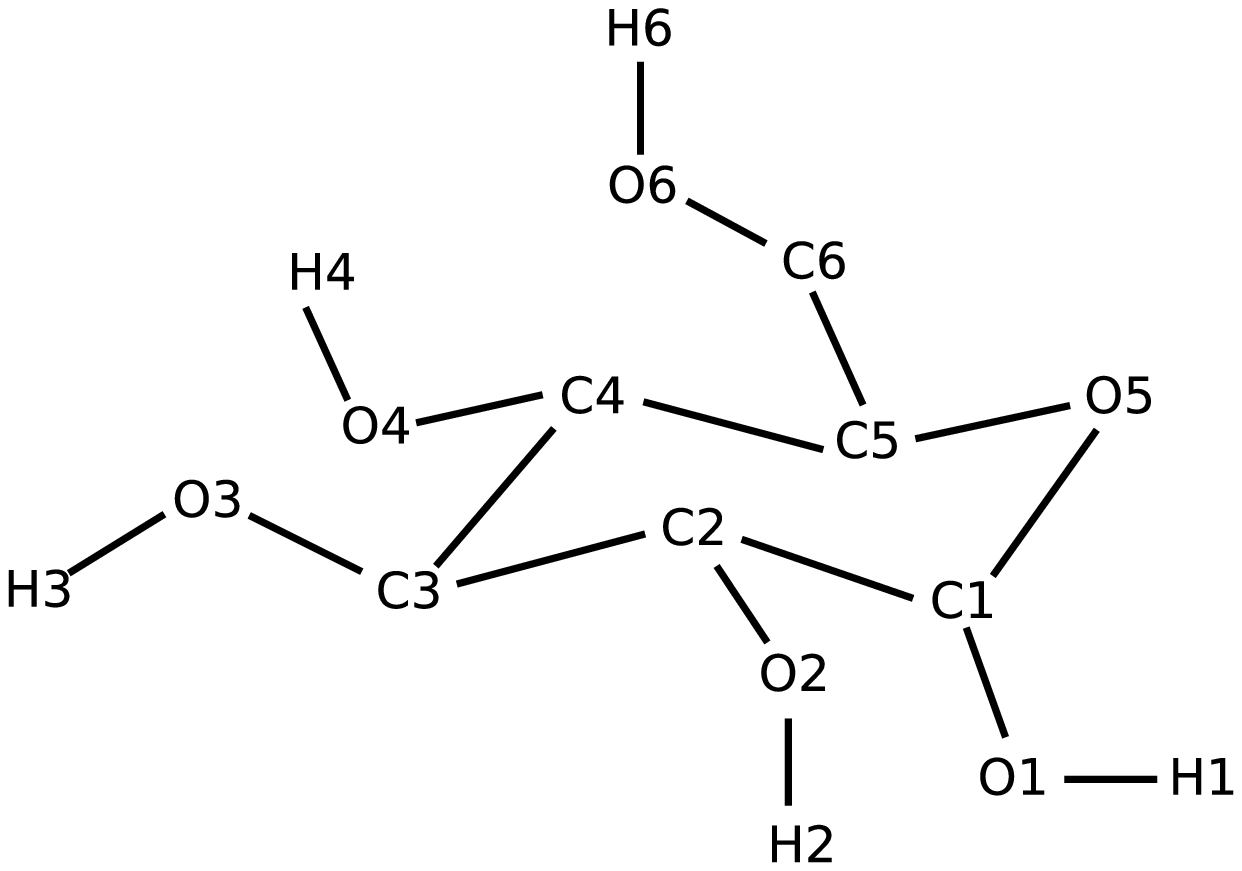}
\caption[Numbering scheme for aldopyranoses rings.]{}
\label{fig:numbering}
\preprintclearpage
\end{figure}

In this work we are focusing on a particular class of carbohydrates, namely,
aldohexoses, saccharides of composition C$_6$H$_{12}$O$_6$. Aldohexoses can
appear in nature in the form of six-membered, cyclic rings, conventionally known as the
pyranose form. Pyranoses are characterized by the presence of five chiral
centers located at the five ring
carbon atoms. Following the convention of Fig.~\ref{fig:numbering}, these chiral carbons are
here denoted as
C1 (the anomeric carbon atom), C2, C3, C4 and C5 (the configurational carbon atom). This leads to quite a high number ($2^5=32$)
of diastereoisomers, characterized by the axial or equatorial orientation of ring substitutents. Based on the chirality at C1 and C5 it is possible to classify pyranoses
into $\alpha$- and $\beta$-pyranoses or \textsc{l}- and  \textsc{d-}pyranoses,
respectively~\cite{McNaught97}. For example, $\alpha$\textsc{-d-}Glc
and $\beta$\textsc{-l-}Glc differ from $\beta$\textsc{-d-}Glc only by the chirality
at C1 and C5, respectively.

For each of these two chiralities, $2^3=8$ stereoisomers remain, which 
are conventionally named~\cite{McNaught97} glucose (Glc), galactose (Gal), mannose (Man), allose (All),
altrose (Alt), talose (Tal), gulose (Gul) and idose (Ido). The position of substituents at
the chiral centres are reported for convenience in Tab.~\ref{tab:chirality}, for $\beta$\textsc{-d-}aldopyranoses.

\begin{table}
\preprintclearpage
\begin{tabular}{lccccc}
\hline
Stereoisomer&\quad{}C1&\quad{}C2&\quad{}C3&\quad{}C4&\quad{}C5\\
\hline
\hline
$\beta$\textsc{-d-}Glc&\quad{}eq&\quad{}eq&\quad{}eq&\quad{}eq&\quad{}eq\\ 
$\beta$\textsc{-d-}Gal&\quad{}eq&\quad{}eq&\quad{}eq&\quad{}ax&\quad{}eq\\
$\beta$\textsc{-d-}Man&\quad{}eq&\quad{}ax&\quad{}eq&\quad{}eq&\quad{}eq\\
$\beta$\textsc{-d-}All&\quad{}eq&\quad{}eq&\quad{}ax&\quad{}eq&\quad{}eq\\
$\beta$\textsc{-d-}Tal&\quad{}eq&\quad{}ax&\quad{}eq&\quad{}ax&\quad{}eq\\
$\beta$\textsc{-d-}Gul&\quad{}eq&\quad{}eq&\quad{}ax&\quad{}ax&\quad{}eq\\
$\beta$\textsc{-d-}Alt&\quad{}eq&\quad{}ax&\quad{}ax&\quad{}eq&\quad{}eq\\
$\beta$\textsc{-d-}Ido&\quad{}eq&\quad{}ax&\quad{}ax&\quad{}ax&\quad{}eq\\
\hline
\end{tabular}
\caption{Orientation of ring substituents (ax: axial; eq: equatorial) for different stereoisomers (referred to the \chair{} conformer).\label{tab:chirality}}
\end{table}

\begin{figure}[t]
\centering
\smallskip
\includegraphics[width=0.98\columnwidth]{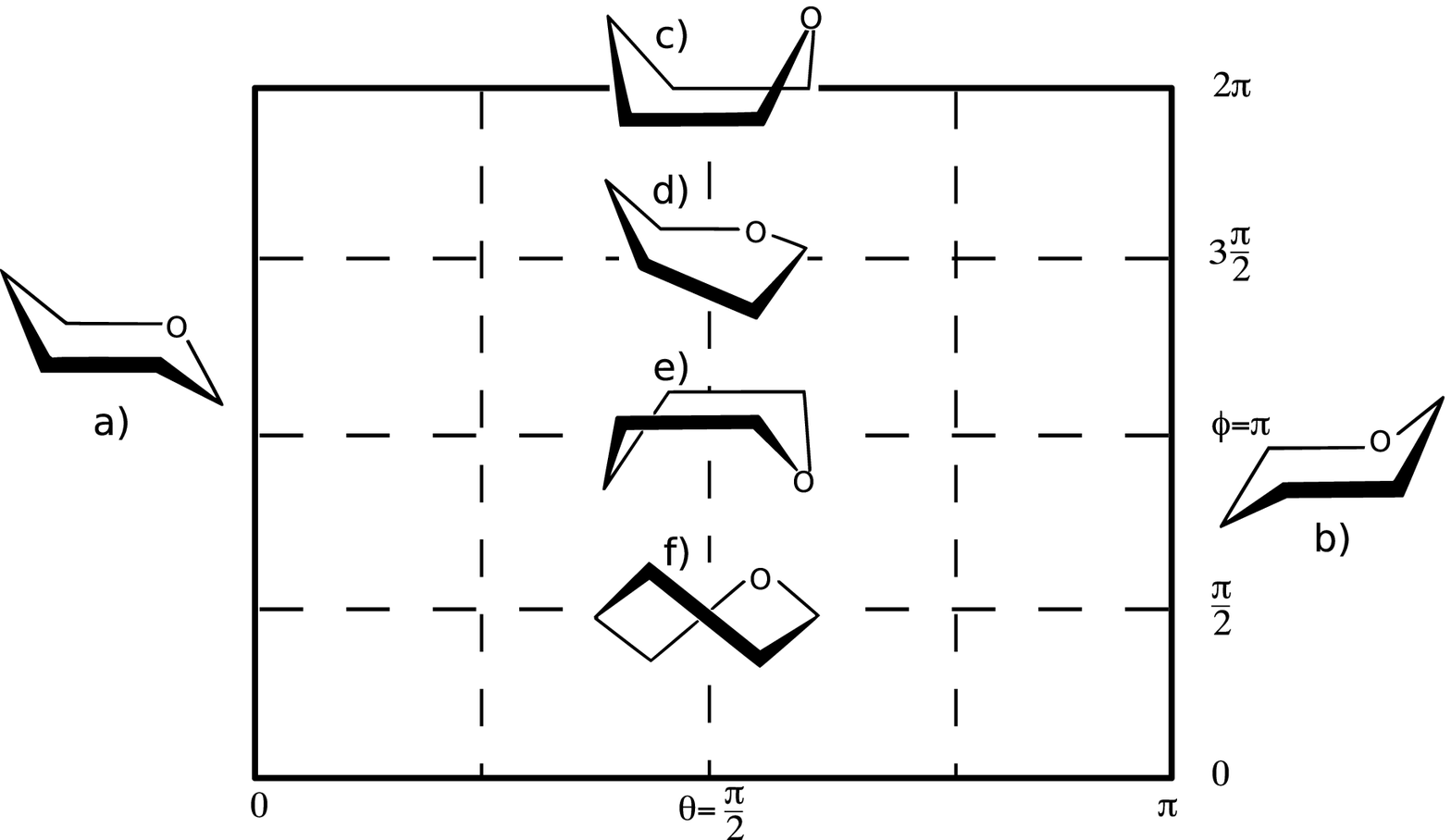}
\caption[Schematic representation of stable ring conformers, located on the
($\theta$,$\phi$) plane of Cremer--Pople coordinates: chair (a) ; inverted
chair (b)  boats (c and e) ; skew boats (e and f). Exact chair
conformers ($\theta=0,\;\theta=\pi$) have been represented out of the
plane, to underline that for them the
$\phi$ angle is not defined.]{\label{fig:conformers}}
\preprintclearpage
\end{figure}

Each of these stereoisomers, in turn, can adopt different ring conformations.
According to IUPAC recommendations~\cite{Rings80}, the conformation
with two parallel ring sides (four coplanar atoms)
and the other two atoms at opposite side of the ring plane is called \emph{chair}.
Chair conformers occur in two forms:
the one with C4 and C1 respectively above and below the plane defined by the other four atoms, 
is identified by the symbol \chair, while the conformer showing C1 above the ring
plane and C4 below it is
denoted by the symbol \invchair{} and often called \emph{inverted chair}.
Other puckered conformers with four coplanar atoms exist,
such as the so-called \emph{boat} and \emph{skew-boat} (the latter also known as twist-boat).
In the case of boats, two parallel ring sides define the ring plane and
the atoms at the extrema of the ring are either above or below the ring plane
(indicated for examples as \boatH{} or \boatL);
in the case of skew-boats the ring plane is defined by three adjacent atoms and the nonadjacent one
with the other ring atoms at opposite side of the ring plane (e.g.
\skew).  Fig.~\ref{fig:conformers} depicts a simplified stereoprojection
of the conformation globe\cite{Cremer95} of a 6-membered ring spanned by the
spherical coordinates $Q, \theta, \phi$ where the puckering amplitude
$Q$ in this work is averaged out during the sampling phase. By
considering $\theta \in (0, \pi)$, the globe surface can be projected
into a rectangular form, which is one of the commonly employed
representation of Cremer--Pople coordinates\cite{Rao98}. The 0 and $\pi$
values of $\theta$ present in Fig.~\ref{fig:conformers} (where the chair
forms are put outside the plane to underline
this) and in subsequent plots are left so that the span of $\theta$ is clear.

Many of the possible conformations are not likely to contribute significantly
to the chemistry and physics of monosaccharides. For pyranoses, only the \chair{} and
\invchair{} conformations happen to be eligible to be the most stable conformations, that is, global
minima which are well separated in energy from neighbouring local ones.
Other conformers such as boats and skew-boats can in principle appear as local,
scarcely populated minima.

The concept of puckering in rings (specifically, in six-membered rings) dates back to the
studies of Sachse\cite{Sachse90,Sachse92} on cyclohexane in the late nineteenth century, and
has been later applied also to saccharidic rings\cite{Sponsler26}. Although it was clear
that the most stable conformers could have been only either of the two chairs, lack of
quantitative information on their relative abundance stimulated many theoretical approaches
(for an historical review the reader can look at Ref.~\onlinecite{Rao98}).  A quantitative
estimate of the free energy difference between puckered conformers in aqueous solution (and
not, as it is sometimes erroneously reported, in vacuum) came first with the analysis performed
by Angyal who assigned, employing an empirical scheme, specific free energy contributions
to different destabilizing interactions. The various terms were derived from measured
equilibria in solution of cyclitols with their borate complexes\cite{angyal69a}. Later,
employing a molecular-mechanical approach, Vijayalakshmi and Rao\cite{Vijayalakshmi72}
obtained other estimates, which were anyway compatible with the inverted chair free energies
predicted by Angyal.

Semiempirical methods have been in general a quite successful approach for determining
the stereochemical properties of numerous cyclic compounds\cite{Barton70}. 
Unfortunately, experimental estimates of the ring conformational
free energy are not feasible for many pyranoses, mostly because the populations of the less stable
conformers are are usually too tiny to be detected with probes such as
NMR\cite{zhu01a,Franks89}. For some
pyranoses, however, the destabilizing effects in the \chair{} and \invchair{} conformations
are similar and, as in the case of idose, the relative population of these two
conformers becomes an experimentally accessible quantity. Indeed, the free
energy differences for idose reported  in
Refs.~\onlinecite{angyal69a} and \onlinecite{Vijayalakshmi72} resulted to be in agreement with the experimental
findings obtained by Snyder and Serianni years later~\cite{Snyder86}. The
agreement with experiment is within $\simeq 4$kJ/mol, but these results are
of particular significance because they correctly predicted the preference of
$\alpha$\textsc{-d-}Ido for the inverted chair conformer.
Except for \textsc{-d-}Alt and, possibly, \textsc{-d-}Gul, only lower bounds
can in principle be determined, given the sensitivity of NMR measurements.
In this sense, anyway, semiempirical methods predict correctly the extremely
high free energy of inverted chairs of glucose, galactose and mannose.  

Some recent experiments involving atomic force microscope (AFM) spectroscopy have allowed
researchers to estimate the puckering free energy of conformers different from the chair
ones. Marszalek, Lee, and coworkers\cite{Marszalek02,Lee04,zhang05a}, for example, estimated
the puckering free energy of glucose twisted boats employing AFM pulling on dextran, cellulose,
and pustulan.  These polysaccharides are all homopolymers of glucose but in the elongation
process, due to the different linkages, cellulose presents (apart from chain flexibility)
only ring deformation, while pustulan presents both ring deformation and rotation around the
C5--C6 bond, but no transition to the twisted boat conformer. By subtracting the free energy
differences related to the various processes, the puckering free energy of the glucose twist
boat was estimated to be about 25 kJ/mol\cite{zhang05a}. It is worth mentioning, that AFM
spectroscopy was first employed to estimate puckering free energies of glucose boats conformers
on carboxymethyl amylose (CMA) \cite{Marszalek98,Li99}. The reported free energy estimates
are in the range 15--18 kJ/mol, therefore quite different form the results obtained from
dextran.  The estimate obtained from CMA, however, could be biased because of the inhability
of a freely jointed chain model to fit the force-extension curves of CMA, which prefers a
pseudo-helical conformation, as it has been pointed out already in Ref.~\onlinecite{Marszalek02}.
Recently, Kuttel and Naidoo\cite{Kuttel05} suggested that in amylose stretching the elongation
might be due to a complex rotation of glycosidic linkages, and that chair to boat transitions
could play little role in the process.

In the field of computer simulations, ab-initio predictions of the puckering free energy
of pyranoses are even more scarce: we are only aware of a Car--Parrinello metadynamics of
glucose in vacuum\cite{Biarnes07} (where, unfortunately, use of non-optimal collective
variables has been made of\cite{Sega09}), a M\o{}ller-Plesset perturbation approach with
inclusion of solvation free energies contributions \cite{Barrows95}, and a DFT calculation
at the $\textrm{B3LYP/6}\textrm{-311++G}^{**}$ level\cite{Appell04}.  A recent, systematic
investigation at the $\textrm{B3LYP/6}\textrm{-311++G}^{**}$ level of all epimers of
glucose\cite{schnupf10a} is worth of notice, although only potential energy has been
calculated.  The calculations of Appell and coworkers\cite{Appell04} predicted a free
energy difference for the \invchair{} conformer of $\alpha$\textsc{-d-}Glc and
$\beta$\textsc{-d-}Glc of 20.40 and  29.18 kJ/mol, respectively.  On the
contrary, the calcultions of Barrows and coworkers\cite{Barrows95}
suggested a free energy of $\beta$\textsc{-d-}Glc \invchair{} conformer
of about 57 kJ/mol.  This estimate appears to be substantially higher than all other
ones reported so far. In absence of similar ab-initio calculations for idose, that could
be employed as benchmark system, it is in our opinion difficult to assess the confidence
level of this estimate.

\section{NMR conformational analysis of  \textsc{d-}altrose\label{sec:nmr}}

The semiempirical and molecular mechanical results suggest that an
estimate of conformer population by means of NMR spectrometry should
be viable also for altrose, which is expected to have a free energy
difference lower than 10 kJ/mol. With the aim of determining this
free energy difference, we performed the complete assignment of the
400 MHz $^1$H spectrum of altrose in water and methanol solution,
also relying on previously published NMR
data\cite{king-morris87a,zhu01a,lichtenthaler97a}, through analysis
of the 1D (differential coupling experiments) and 2D (mainly HSQC)-NMR
techniques.  Gradient enhanced version of the DQF-COSY, TOCSY, HSQC
and HMBC techniques were used for signal assignments whereas
differential $^1$H/$^1$H decoupling, J-resolved spectroscopy  and
high resolution HSQC measurements were used
for establishing the proton J couplings.

The $^1$H-NMR analysis of \textsc{d}-altrose is strongly hindered
by the huge overlap among $^1$H-signals of the different structural
isomers (pyranose and furanose forms) among which it distributed
in aqueous solution. The determination of the $^3$J$_\textrm{HH}$
coupling patterns of the altropyranose, as presented in
Tab.~\ref{tab:coupling}, mainly derive from the high digital
resolution (4096$\times$1024 datapoints) HSQC-NMR spectrum.

\begin{table*}
\begin{small}
\begin{tabular}{c | cclccll | cclccll}
&\multicolumn{6}{c}{$\alpha$\textsc{-d-}\textbf{altropyranose}}&&     \multicolumn{6}{c}{$\beta$\textsc{-d-}\textbf{altropyranose}}&\\
H/C Type &  $\delta_H^a$&  $\delta_C^a$ &  {$\!\!\!^3$J$_{HH}^a$}& $\delta_H^b$ &  $\delta_C^b$&  {$\!\!\!^3$J$_{HH}^b$} && $\delta_H^a$ &  $\delta_C^a$ &  {$\!\!\!^3$J$_{HH}^a$} & $\delta_H^b$ &  $\delta_C^b$&  {$\!\!\!^3$J$_{HH}^b$} &\\
\hline
\hline
H$_1$/C$_1$    &$\;5.01\;$  & $\;95.3\;$  &$\;$3.4+\Wc &4.93&96.3 &1.5+\Wc&& 5.14 &93.3  &1.4+\Wcc&5.04&93.5 &1.3+\Wcc\\
H$_2$/C$_2$    &3.87        &71.9         &3.4, 5.6              &3.81&71.4&1.5, 3.6     &&                       3.85 &72.3  &1.4, 4.1       & 3.69 &73.1 & 1.3, 3.9&   \\
H$_3$/C$_3$    &3.95        &71.8         &3.7, 5.6              &3.88&72.9&3.0, 3.6     &&                       3.83 &75.6  &2.2, 4.1       & 3.95 &72.7 & 3.0, 3.9    \\
H$_4$/C$_4$    &3.91        &66.8         &3.7, 7.6              &3.80&65.6&3.0, 9.8     &&                       3.84 &65.6  &2.2, 9.1        & 3.74 &66.0 & 3.0, 9.8    \\
H$_5$/C$_5$    &4.09        &72.8         &3.4, 6.3, 7.4         &3.98&70.5&2.7, 5.5, 9.8&&                       4.08 &72.1  &2.2, 4.9, 9.1  & 3.70 &75.8 & 2.0, 5.5, 9.8\\
H$_6$/C$_6$    &3.81        &62.1         &6.3, 12.0             &3.75&63.2&5.5, 11.8    &&                       3.75 &63.1  &2.2, 11.7      & 3.67 &63.6 & 5.5, 11.7   \\
H$_{6'}$/C$_6$ &3.83        &             &3.4, 12.0             &3.83&    &2.7, 11.8    &&                       3.90 &      &4.9, 11.7      & 3.81 &     & 2.0, 11.7   \\
\hline
\hline
\multicolumn{14}{l}{Chemical shifts of $^{13}$C ($\delta_C$)
and $^1$H ($\delta_H$) are reported in ppm;  coupling constants are reported in Hz.}\\
\multicolumn{14}{l}{$^a$ in D$_2$O; $^b$ in CD$_3$OD. }\\

\end{tabular}
\end{small}
\caption{Assigned chemical shifts and coupling constants for the four main forms of altropyranose.}
\label{tab:coupling}
\end{table*}

We have chosen J$_{\textrm{2-3}}$ (ax-ax in $^1$C$_4$)  and
J$_{\textrm{4-5}}$ (eq-eq in $^1$C$_4$) as sensitive probes for the
conformational transition. Since we were expecting altrose to exists
almost only in $^4$C$_1$ conformation in solvents with lower dieletric
permittivity, we have recorded and assigned the spectra also for
altrose in deuterated methanol, thus considering its coupling
patterns to be representative of the limit forms in water,
J(ax/ax)=9.8 and J(eq/eq)=3.6. A subsequent metadynamics simulation
of altrose in methanol employing the force field presented in this work
(Section~\ref{sec:reparam}) showed indeed a dramatic bias towards the
$^4$C$_1$ form with respect to the results in water, confirming the
trend we expected.  Using these values of the coupling constants
for the limit forms, the measured J$_\textrm{2-3}$ and J$_\textrm{4-5}$
in D$_2$O appear to be the average from a population of 68 and
65\% $^4$C$_1$, respectively.
Thus, our analysis leads to an estimated molar fraction for the $^4$C$_1$
conformer of
0.66$\pm$0.02. The same procedure can be employed to estimate the molar
fraction
of the puckered conformers of  $\beta$\textsc{-d-}altrose, assuming
an averaged J(eq/eq) = 3.6 Hz for all $^{4}$C$_{1}$ conformations and an
averaged J(ax/ax) = 9.8 Hz for all $^{1}$C$_{4}$ conformations. 
This leads to an estimated molar fraction of 0.90 $\pm$ 0.02 for the $^4C_1$
conformer.

It has to be noted that a decade ago, Lichetenthaler and
coworkers\cite{immel99a,lichtenthaler97a,Nogami97a} 
found that $\alpha$-cycloaltrin could exist as a fast equilibrating mixture of three
different conformations (\chair,\invchair{} and $^\mathrm{O}$S$_2$) 
in almost the same relative molar ratio. If we apply the Lichtenthaler three states model to the
monosaccharide $\alpha$ altroppyranose, the relative population of the
$^4$C$_1$, $^\mathrm{O}$S$_2$ and $^1$C$_4$ would be  62:19:19.
Even Snyder and Serianni discussed the possibility of that skew conformation
(reported with the alternative name of $^3$S$_5$) being a possible form of $\alpha$\textsc{-d-}idose.
However, their conclusion was based on the fact that the measured
$^3$J$_{H4,H5}$ was not fully compatible with the inverted chair
form. In our case, certainly, nothing prevents to employ a three
state model but, differently from the case of idose,  there is also
no explicit need to introduce a population of skwes to interprete the
averaged coupling constants.

By inverting Boltzmann formula under the two-states assumption, one can estimate the free
energy of the inverted chair to be 1.6$\pm$0.2 and 5.5$\pm$0.5 kJ/mol for $\alpha$ and $
\beta$\textsc{-d-}altrose, respectively. The theoretical estimates of Angyal\cite{angyal69a}
(0.8 and 8.4 kJ/mol) and of Vijayalakshmi and Rao\cite{Vijayalakshmi72} (4.4 and 8.9 kJ/mol
) are thus in agreement with our measured values within roughly 3 kJ/mol.
By and large, these results on altrose confirm
that the estimates of Refs.~\onlinecite{angyal69a} and \onlinecite{Vijayalakshmi72} represent
reasonable approximations of pyranoses free energy difference between chair
conformers. Thus, in this work we will use these theoretical estimates as a 
reference when testing new force field parameter sets.

\section{Computational Methods\label{sec:methods}}
\subsection{Refining Metadynamics with Umbrella Sampling}
The term metadynamics identifies a number of techniques that have been devised during the
last decade to accelerate dynamics for systems displaying meta-stabilities (see for example Refs.~\onlinecite{Laio02,Iannuzzi03,Laio05a,Bussi06}).
All these variants share the usage of a time-dependent biasing
potential, $U_{\rm{}bias}[s(x),t]$, to ease the exploration of the phase space along
suitably chosen collective variables, $s(x)$, in turn themselves functions of the atomic coordinates $x$. 
The collective variables must represent all
slow degrees of freedom characterizing the system, in order for metadynamics
 -- as well as any other free energy profile reconstruction method -- to be
meaningful\cite{Laio08}.

The direct metadynamics approach\cite{Laio05a} shares many traits with other enhanced sampling methods,
such as the Local Elevation method\cite{Huber94} or the adaptive umbrella
sampling\cite{Mezei87} (detailed lists of enhanced
sampling techniques based on molecular dynamics can be found in Refs.~\onlinecite{hansen10a}
and \onlinecite{Laio08}), but the fact that the free energy landscape in metadynamics is
usually estimated as the negative of the biasing potential in a single, out-of-equilibrium
sweep, makes the technique susceptible, at least in principle, of errors introduced by the deposition
protocol. Distinct approaches have been devised to estimate the statistical and
systematic errors in 
metadynamics, and also to recover a truly equilibrium free energy
profile\cite{Laio05a,Bonomi09,Babin06}. Among these,
the approach proposed by Babin and coworkers\cite{Babin06} is in our opinion one of the most intuitive and
versatile ones, since it allows in a simple way to simultaneously estimate the statistical error
and to eliminate systematic errors introduced by the deposition protocol. The basic idea
is the following: a metadynamics run is performed up to the build-up time $t_b$, so that the whole collective variables
space has been explored, and the total potential energy -- sum of the physical and the
bias potential -- at the end of the run reads
\begin{equation}
V(x)=U(x)+U_{\mathrm{bias}}[s(x),t_b] \quad.
\end{equation}
At this point, a molecular dynamics simulation in the potential $V$ is performed, much in
the spirit of umbrella sampling, whereas the biasing potential has been determined in an
adaptive way by the metadynamics run. The dynamics is 
characterized by an almost diffusive behavior in the collective variables space, since the
meta-stabilities have been removed by the bias potential, provided that all the states of
the new potential $V(x)$ are separated by
energy barriers comparable or lower than the thermal energy scale $k_\textsc{b}T$.
The deviation from a truly diffusive behavior is due by 
residual features of the free energy landscape which are originated from the statistical and systematic
errors in the metadynamics run. During this 
equilibrium run, the phase space is sampled with probability density
\begin{equation}
\rho_\mathrm{bias}(s)=\frac{e^{-\beta
\{ F(s)+U_\mathrm{bias}[s(x),t_b] \}}}{\int \mathrm{d}s e^{-\beta\{
F(s) +U_\mathrm{bias}[s(x),t_b]\}}}
\end{equation}
that can be estimated by computing the histogram
$\rho_\mathrm{bias}(s)=(t-t_b)^{-1}\int_{t_b}^t{\delta[s-s(t^\prime)] dt^\prime }$ during the run.
Eventually, the free energy profile is given, up to an additive constant, by
\begin{equation}
F(s)=-U_\mathrm{bias}[s(x),t_b]-k_{\textsc{b}}T \ln \rho_\mathrm{bias}(s).\label{eq:master}
\end{equation}
In the exact expression Eq.~\ref{eq:master}, the term $k_{\textsc{b}}T \ln \rho_\mathrm{bias}(s)$ can be regarded as a
correction to what is the standard metadynamics estimate of the free energy landscape,
$F_\mathrm{meta}(s)= -U_\mathrm{bias}(s)$. This correction term compensates for the
systematic errors introduced by the deposition protocol, which are not completely  under control in
the metadynamics run. In other words, from a simple metadynamics run there is no way to
guarantee that the term $\rho_\mathrm{bias}(s)$ has become a constant within statistical
fluctuations.
The corresponding statistical error can be estimated from the fluctuations of $\rho_\mathrm{bias}(s)$ using standard
error analysis. Differently from the approach involving only a metadynamics run, one should
not worry about the speed of the deposition process (which involves the height, width and
deposition rate of the Gaussian functions) as long as the subsequent equilibrium run is ergodic.
This means that the Gaussian functions placed during the build-up phase have only to (a) be smaller than
$k_\textsc{b}T$, (b) being of width comparable or smaller than the finest detail of the free energy
landscape which is \emph{deeper} than $k_\textsc{b}T$ and (c) cover the whole conformational space
of the collective variables.

This idea of employing the biasing potential with an umbrella-like sampling has been
applied, besides to metadynamics\cite{Babin06}, also to the local elevation
method\cite{hansen10a}. 

\subsection{Puckering Coordinates\label{subsec:CPcoord}}
The generalized coordinates introduced by Cremer and Pople\cite{Cremer75,Cremer90} can be 
used to identify puckered conformations of rings with an arbitrary number of members. Their
definition makes use of the projections $z_j$ of the position vector of each ring atom
onto the normal of the mean ring plane.
In the case of six-membered rings the Cremer-Pople coordinates can be defined in terms
of the distances $z_j$ as a functions of 3 parameters $q_2$, $\phi_2$ and $q_3)$, by
\begin{equation}
q_2 \cos \phi_2 =  \sqrt{\frac{1}{3}} \sum_{j=1}^6 z_j \cos\left[\frac{2\pi}{3}(j-1)\right] \label{eq:qcosphi}
\end{equation}
\begin{equation}
q_2 \sin \phi_2 = -\sqrt{\frac{1}{3}} \sum_{j=1}^6 z_j \sin\left[\frac{2\pi}{3}(j-1)\right] \label{eq:qsinphi}
\end{equation}
\begin{equation}
q_3 = \sqrt{\frac{1}{6}} \sum_{j=1}^6 (-1)^{j-1} z_j \label{eq:qeven} \quad .
\end{equation}
This coordinate set $(q_2,\phi_2,q_3)$ can be conveniently expressed as a spherical coordinate set $(Q,\theta,\phi)$,
\begin{equation} 
\left\{ 
\begin{array}{l} 
q_2\cos\phi_2=Q\sin\theta\cos\phi \\
q_2\sin\phi_2=Q\sin\theta\sin\phi \\
q_3=Q\cos\theta, 
\end{array}
\right.
\end{equation}
where $\theta\in[0,\pi]$, $\phi\in[0,2\pi)$ and $Q$, the so-called total puckering
amplitude, is defined as  $Q^2\equiv\sum_{j=1}^N z_j^2=\sum_m q^2_m.$

There are two main interconversion paths in pyranoses, namely, the inversion path --- connecting
the \chair\ and \invchair\ conformers --- and the 
pseudo-rotation path\cite{Cremer75} --- connecting the more flexible boat and skew-boat
conformers --- that can be easily represented in this coordinate set. The inversion path develops along the
$\theta$ coordinate from the \chair\ conformer at $\theta=0$ to the \invchair\ one at
$\theta=\pi$, while the pseudo-rotation one develops along $\phi$, at $\theta=\pi/2$.
Notice that $\phi$ is a $2\pi-$periodic coordinate,  meaning that points at $\phi=0$ or
$\phi=2\pi$ (at a given value of $\theta$) represent the same conformer. This is not true
for $\theta$, which is not periodic.

In this sense, the spherical set ($Q,\theta,\phi$) has the advantage that only the two angular variables
are needed as collective variables in order to perform a proper -- that is, ergodic and unbiased -- exploration
of the puckered conformations space of typical six-membered rings. This is because along
the radial direction no meta-stabilities occur, and the radial coordinate $Q$ relaxes fast enough to be ergodic
for every reasonable set of potential (that is, when the bond lengths are rigid or
quasi-rigid).
As we showed in a previous work\cite{Sega09}, not every representation
of the Cremer--Pople coordinates is equivalent to the end of being used as collective variables for a
conformational search. In particular, any two-dimensional subset of Cremer--Pople coordinates
whose functional form involves also biasing forces along the direction of $Q$ might suffer from
lack of ergodicity and, therefore, lead to biased sampling\cite{Laio08}. In this work we
make only use of the angular variables of the spherical coordinate set as collective
variables for the metadynamics run.

Alternative generalized coordinates can be in principle employed to characterize the puckered
conformers of six-membered rings, such as the three out-of-plane dihedrals introduced by
Strauss and Pickett\cite{Strauss70}, or other definitions based on three internal dihedral
angles\cite{Zefirov90,Haasnoot92,Berces01}. All these alternative schemes produce good puckering
coordinates, in the sense that they allow to uniquely map the complete puckering conformational
space. However, in the context of conformational search using accelerated methods, oppositely
to Cremer--Pople coordinates, (a) they require the exploration of the complete three-dimensional
space, being therefore less convenient than a two dimensional phase space search, and (b)
the set of micro-states corresponding to the different conformers generally define less
simple surfaces which, as it will be discussed later, might affect the determination
of the relative populations of conformers.

\subsection{Simulation Details}
The metadynamics and equilibrium simulations have been performed using a version of the {\sc
grometa} simulation package\cite{Camilloni08,Berendsen95,Lindahl01,Spoel05}, previously
modified to allow the usage of the $\theta$ and $\phi$ angular coordinates of Cremer and
Pople, as described in Ref.~\onlinecite{Sega09}.  For each of the 16  \textsc{d-}aldopyranoses,
a system composed of the respective sugar ring in a cubic simulation box filled with 504
water molecule was set up.  The SPC\cite{Berendsen84} model has been used to describe the
water molecules employed to solvate each of the pyranoses. The {\sc settle} algorithm has
been used to make water rigid\cite{Miyamoto92}, and all bond lengths in the sugar molecules
have been constrained using the \textsc{shake} algorithm\cite{Ryckaert77}. An integration
step of 0.2 fs has been used for every phase of the simulations. Before starting each run,
a 100 ps long molecular dynamics simulation with no bias has been performed to equilibrate
the different sugars in their \chair{} conformer.

The metadynamics part of the run consisted in a 4 ns long simulation, using Gaussian potential
functions of height 0.5 kJ/mol and with 0.05 rad for both angular variables, to build up the
biasing potential.  Gaussian functions have been placed every 200 integration steps.  The
Nos\'e--Hoover\cite{Nose84,Hoover85} thermostat and the Parrinello--Rahman\cite{Parrinello81}
barostat have been applied to simulate isothermal-isobaric conditions at 300 K and 1 atm
using relaxation times of 0.1 and 1.0 ps, respectively. The simulation box edges have been
kept orthogonal, and have been rescaled using an isotropic pressure coupling, which controlled
the trace of the pressure tensor.

Each metadynamics run has been then followed by a 4 ns long equilibrium molecular dynamics run at 
the same thermodynamic conditions, using the set of Gaussians
placed during the metadynamics to generate the time-independent biasing potential. At a
difference with Ref.~\onlinecite{Babin06}, where the Lagrangian metadynamics with truncated
Gaussians has been employed, we make use of the standard direct metadynamics. During this
run, the histogram $\rho_\mathrm{bias}(\theta,\phi)$ has been collected by sampling configurations every
40 fs on a grid of $60\times60$ points.

\section{Puckering Free Energy of Glucose\label{sec:results}}
\begin{figure}[t]
\centering
\smallskip
\includegraphics[width=1.\columnwidth]{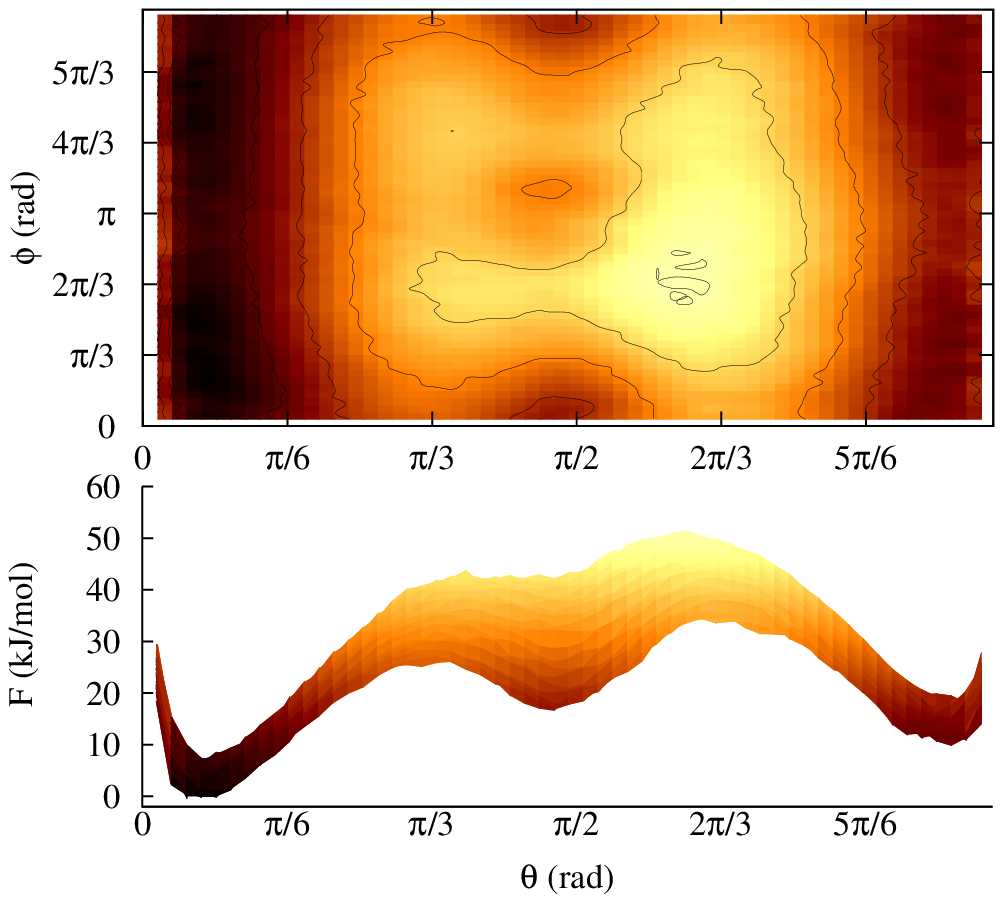}
\caption[Puckering free energy landscape (already corrected with the results from the umbrella sampling phase) 
of $\beta$\textsc{-d-}Glc using the standard 45a4
parameter set. Upper panel: the $(\theta,\phi)$ plane is shown, with isolines drawn
every 10 kJ/mol starting from the absolute minimum. Lower panel: projection of the free
energy profile onto the $\phi=0$ plane. Darker colors correspond to lower energies.]{\label{fig:GlcStd}}
\end{figure}

\begin{figure}[t]
\centering
\smallskip
\includegraphics[width=1.\columnwidth]{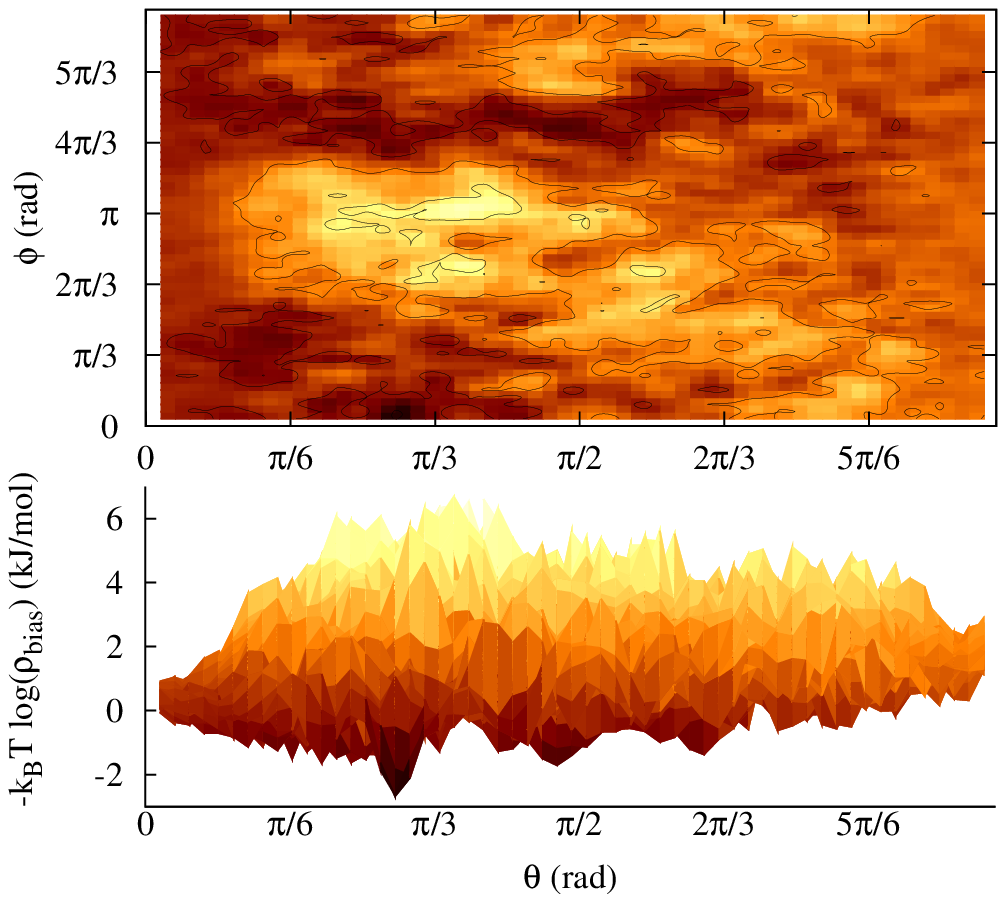}
\caption[Contribution to the total puckering free energy of  $\beta$\textsc{-d-}Glc (45a4 parameters) 
from the equilibrium sampling, $-k_\textsc{b}T\ln{\rho_\mathrm{bias}}(\theta,\phi)$. An immaterial
constant has been added, so that the correction term is zero at the position of the
free energy global minimum. Isolines are drawn every 2.5 kJ/mol (upper panel). 
The projection onto the $\phi=0$ plane of the correction term (lower panel).]{\label{fig:corr_GlcStd}}
\end{figure}

The combined metadynamics--umbrella sampling has been first applied to the calculation of the
puckering free energy profile of $\beta$\textsc{-d-}Glc, employing the standard
\textsc{gromos} 45a4 parameter set\cite{Lins05}. In Fig.~\ref{fig:GlcStd} we present the free
energy profile as a function of the Cremer--Pople angular variables $\theta$ and $\phi$,
showing isolines separated by 10 kJ/mol (upper panel). In order to facilitate the comprehension
of the plot, the projection of the free energy profile onto the $\phi=0$ plane (lower panel)
is also provided: the lower contour of the colored region in the projected profile allows
to easily identify the minima along the $\theta$ coordinate (the minima of
this lower contour) and the transition states, or free energy saddle points
(the maxima of the lower contour). On the
$(\theta,\phi)$ plane, \chair{} (0,-), \invchair{} ($\pi$,-), and  \boatH{} ($\pi/2$,0)
conformers are clearly recognizable as minima basins.  Notice that, due to the periodic nature of $\phi$, the \boatH{}
conformer appears in this representation to be split in two across the $\phi=0$ line.
Another local minimum basin, more shallow than the previous ones is located around the
$^1$S$_3$ ($\pi/2$,$7\pi/6$) conformer and seems to include
also some other near boat-like conformer. This kind of occurrence can be a natural feature for gluco-pyranoses,
because of the high flexibility of the ring conformers along the pseudo-rotational path
(located at $\theta=\pi/2$). A more detailed description of this local minimum is anyway beyond
the interest of this work, because of its very high free energy.

In addition to the free energy profile, in Fig.~\ref{fig:corr_GlcStd} we show an analogous
plot, presented both on the $(\theta,\phi)$ plane (upper panel), and projected onto the
$\phi=0$ plane (lower panel), of the correction to the free energy profile coming from the
umbrella sampling phase, $-k_{\textsc{b}}T \ln \rho_\mathrm{bias}(s)$. Performing a screening of this
contribution is quite important, because it allows to check if residual meta-stabilities
(recognizable as basins) are left after the metadynamics phase, and whether an ergodic sample
of the interesting region in the $(\theta,\phi)$ plane has been performed or not: poorly
sampled regions will appear as peaks. In Fig.~\ref{fig:corr_GlcStd} it is possible
to see that the metadynamics run performed reasonably well. Indeed, the paths connecting
the original metastable states do not show residual metastabilities. It is seen that the
complete space of puckering angles has been sampled properly. Corrections to the free energy
difference between \chair{} and \invchair{} are $\simeq 2$~kJ/mol. In case of the next most
populated state, \boatH{}, the correction increases to $\simeq4$~kJ/mol, while for other
states the correction can reach, but never exceeds, $\simeq 6$~kJ/mol with respect to the
free energy of \chair{}. The umbrella phase thus contributes in a significant way to the
estimate of free energy differences, and --- as expected --- appears to be more important
for less populated states and for transition states.

Both the \invchair{} and \boatH{} puckering free energies (local minima at about 10 and 17
kJ/mol, respectively)  appear to be lower than what one would expect. In particular, the
free energy of the inverted chair is about 15 kJ/mol lower than both our reference
values\cite{angyal69a,Vijayalakshmi72}.  This free energy difference leads to a inverted
chair population of about 1\%, which can be observed during regular molecular dynamics runs
at equilibrium.  While not strictly incompatible with NMR experiments, which usually
can predict populations with a $\simeq$2\% accuracy, this value is certainly 
strikingly lower than all other theoretical
estimates. This fact is even more important if one considers that $\beta$\textsc{-d-}Glc
is supposed to have the largest \invchair{} free energy among aldopyranoses. It is thus expected
that the \invchair{} conformers of the other stereoisomers could be characterized by even lower free energy differences. 
As it will be discussed in Sec.\ref{sec:parameters}, this scenario is only
partially correct, as unexpected patterns of the chair--inverted chair free energy
difference appear along the series.

A more direct connection with experimentally measurable quantities is performed through the calculation
of the population of the basins associated with each of the recognizable conformers. To this
aim, the $(\theta,\phi)$ plane has been partitioned in four regions covering the four main
basins:
$\theta\in[0,\pi/3]$, associated to \chair;
$\theta\in [2\pi/3,\pi]$, associated to \invchair;
$\theta\in[\pi/3,2\pi/3]$ and $\phi\in [-2\pi/3,2\pi/3]$, associated to \boatH;
$\theta\in[\pi/3,2\pi/3]$ and $\phi\in[2\pi/3,4\pi/3]$, associated to the mixture around $^1$S$_3$.
This partitioning of the conformational space appears to be quite natural, since the separating
lines are located with good approximation along the maxima of the free energy surface (as
is evident especially from the side-view of the free energy landscape, lower panel of 
Fig.~\ref{fig:GlcStd}) and, therefore, also close to the transition states.
The population of a region $\Omega$ can then be calculated simply as 
\begin{equation}
p(\Omega)=Z^{-1}\int_{\Omega} e^{-F(\theta,\phi)/k_\textsc{b}T} d\theta d\phi 
\end{equation}
where the integral of the normalization factor 
\begin{equation}
Z=\int e^{-F(\theta,\phi)/k_\textsc{b}T} d\theta d\phi 
\end{equation} is extended to the whole $(\theta,\phi)$ space.
According to this numerical estimate, the \chair,\invchair,\boatH{} and $^1$S$_3$ conformations
account for 98.42, 1.52, 0.0461 and 3.7$\times10^{-4}$\% of the total population, respectively.

When the manuscript was in preparation, we discovered that Hansen and H\"uneneberger already
performed an analysis of the puckering free energy of glucose\cite{hansen10a}. To our surprise,
their estimates of the inverted chair free energy differ by a non-negligible amount from
that presented here, although the same force field and thermodynamic conditions have been
used.  In the approach of Hansen and H\"unenberger, free energies
are derived from populations by inverting Boltzmann formula. We did the same and obtained a
value of 10.43~kJ/mol for  the free energy of \invchair{}, while in Ref.~\onlinecite{hansen10a}
this value ranged from 16.0 to 16.5~kJ/mol (see Tab.4 in Ref.~\cite{hansen10a}) The difference
between these two approaches is quite important, as it translates to inverted chair populations
of $\simeq 1\%$ and $\simeq 0.1\%$ for the results of this work and of Ref.\onlinecite{hansen10a},
respectively. Notice that in a recent work Spiwok and coworkers found the \invchair{} free
energy of $\beta$\textrm{-d-}Glc  to be 11.6$\pm$1.8 (thus, compatible with our findings)
using the 45a4 force field and a metadynamics employing all three Cremer-Pople coordinates
as collective variables.

We give here a tentative explanation of the difference with the results of
Ref.~\onlinecite{hansen10a}.
The difference is most probably not related to which set of
collective variables (Pickett dihedrals versus Cremer--Pople coordinates) has been used to
enhance the sampling, because both sets of coordinates represent every slow degree of freedom
and allow to span the three-dimensional puckering conformational space in an ergodic way.
Systematic errors related to the accelerated dynamics should have been eliminated in both
approaches, since both sampling methods provide an unbiased estimate, thanks to the equilibrium
sampling. Hence, among the possible reasons left are: the use of different algorithms for the simulation of
isothermal--isobaric conditions;  the different system size;  cut-off and long range corrections.
Another explanation could the way different states are defined:
partitioning a three dimensional conformational space is far more complicate than partitioning
a two dimentional space. In the former case identify in a proper way the different basins
might lead to a miscounting of states, while in the latter
one only straight cuts along the $\theta$ and $\phi$ directions are needed.
Still, more investigation would be needed to address properly this subject, which is out of
the scope of this work.

\section{Puckering Free Energy of Aldopyranoses\label{sec:parameters}}
\subsection{The 45a4 Parameter Set}
Apart from the problems related to a correct definition of different puckered states,
it is clear that the calculation of the free energy profile of $\beta$\textsc{-d-}Glc
alone cannot be
considered exhaustive nor satisfactorily. The theoretical estimates predict a great variety in the 
conformational free energies of the 16 stereoisomers of $\beta$\textsc{-d-}Glc, and puckering properties have
to be checked separately for each of them. For these reasons
we decided to compute in a systematic way the puckering free energy landscape for the whole series of $\alpha$
and $\beta$\textsc{-d-}pyranoses. 

From the point of view of the force field, the
stereoisomers of glucose differ only slightly, namely (a) the order of the two central atoms
involved in an improper dihedral interaction at a chiral
centre has to be inverted, in order to move a residue form the equatorial to the axial
conformation; (b) $\alpha$ anomers are distinguished from $\beta$ anomers by
different torsional interactions on the O5--C1--O1--H1 dihedral, and (c) those sugars having O4 and C6
located on the same side of the ring plane (galactose,  talose, gulose,  idose)
are modeled with different parameters for the O5--C5--C6--O6 and C4--C5--C6--O6 dihedral angles\cite{Lins05}.
\begin{table*}
\begin{tabular}{l|r@{.}l@{$\,\pm\,$}r@{.}lr@{}lr@{.}l@{$\,\pm\,$}l@{.}lr@{,}lr@{.}r@{$\,\pm\,$}r@{.}lc|r@{.}lr@{.}lr@{.}l@{}l}
\hline
Isomer & \multicolumn{4}{c}{$\Delta{}G$ [\invchair{}]$^a$}&\multicolumn{2}{c}{Next$^b$}
&\multicolumn{4}{r}{$\Delta{}G$[Next]$^c$ } & \multicolumn{2}{c}{$\textrm{TS}^d$}&
\multicolumn{4}{r}{$\Delta{}G$[TS]$^e$ } &&\multicolumn{2}{c}{P[\chair{}]$^f$ }&
\multicolumn{2}{c}{P[\invchair{}]$^g$} & \multicolumn{3}{c}{P[Next]$^h$ } \\ 
 \hline 
 \hline 
 
$\beta$\textsc{-d-}Glc	&     10&0 & 0&2   & $^{3\mathrm{O}}$&B  &     16&7 & 0&2   	&(1.06&0.10)  &25&9 & 0&2&&   98&42(1) &         1&52(1)   	&  4&6(1)&$\times10^{-2}$ \\
$\beta$\textsc{-d-}Gal	&     4&6 & 0&2    & $^3$&S$_1$    	 &     21&5 & 0&2   	&(1.11&3.94)  &38&6 & 0&4&&     91&2(1)  &  	    8&8(1)      &  6&2(1) &$\times10^{-3}$ \\
$\beta$\textsc{-d-}Man	&     3&0 & 0&2    & $^\mathrm{O}$&S$_2$ &     23&2 & 0&2   	&(1.11&2.44)  &41&2 & 0&2&&     80&6(1)  &  	   19&4(1)      &  4&3(1) &$\times10^{-3}$ \\
$\beta$\textsc{-d-}All	&     21&2 & 0&2   & $^{3\mathrm{O}}$&B  &     27&9 & 0&2   	&(1.17&3.40)  &36&9 & 0&3&&     99&98(1)&  	    0&0205(2)   &  4&6(1) &$\times10^{-4}$ \\
$\beta$\textsc{-d-}Tal	&    -2&8 & 0&2    & $^{3\mathrm{O}}$&B  &     36&8 & 0&2   	&(1.11&3.62)  &47&6 & 0&2&&     58&7(2)  &  	   41&3(1)      &  3&8(1) &$\times10^{-5}$ \\
$\beta$\textsc{-d-}Gul	&     14&8 & 0&3   & $^\mathrm{O}$&S$_2$ &     33&9 & 0&1   	&(1.22&3.30)  &43&2 & 0&2&&     99&92(1)&  	    0&08(1)     &  5&7(1) &$\times10^{-5}$ \\
$\beta$\textsc{-d-}Alt	&     15&5 & 0&2   & $^\mathrm{O}$&S$_2$ &     33&7 & 0&1   	&(1.11&3.94)  &42&4 & 0&2&&     99&87(1)&  	    0&129(2)    &  9&4(1) &$\times10^{-5}$ \\
$\beta$\textsc{-d-}Ido	&     13&0 & 0&2   & &B$_{25}$      	 &     35&0 & 0&3   	&(1.06&4.47)  &44&9 & 0&1&&     99&59(1)&  	    0&412(4)    &  6&1(1) &$\times10^{-5}$ \\
\hline\hline
$\alpha$\textsc{-d-}Glc	&     3&7 & 0&2    & $^\mathrm{O}$&S$_2$ &     25&7 & 0&2   	&(1.11&0.21)  &35&2 & 0&2&&     90&85(6)        &    9&15(7)    &  1&7(1)&$\times10^{-3}$ \\
$\alpha$\textsc{-d-}Gal	&     2&1 & 0&2    & $^3$&S$_1$    	 &     33&0 & 0&2   	&(1.11&3.40)  &47&1 & 0&2&&     74&9(2)    	&    25&1(2)    &  5&8(1) &$\times10^{-5}$ \\
$\alpha$\textsc{-d-}Man	&     2&3 & 0&3    & $^\mathrm{O}$&S$_2$ &     15&2 & 0&3   	&(1.11&3.30)  &29&5 & 0&4&&     71&9(3)    	&    28&0(2)    &  3&9(1)&$\times10^{-2}$ \\
$\alpha$\textsc{-d-}All	&     18&2 & 0&2   & $^{3\mathrm{O}}$&B  &     35&6 & 0&2   	&(1.65&4.79)  &56&6 & 0&3&&     99&94(1)  	&    0&060(1)   &  3&4(1) &$\times10^{-5}$ \\
$\alpha$\textsc{-d-}Tal	&     1&2 & 0&3    & $^1$&S$_3$    	 &     32&1 & 0&3   	&(1.11&3.51)  &37&1 & 0&3&&     63&7(4)    	&    36&3(3)    &  3&0(1) &$\times10^{-4}$ \\
$\alpha$\textsc{-d-}Gul	&     12&9 & 0&2   & $^\mathrm{O}$&S$_2$ &     42&7 & 0&2   	&(1.06&3.72)  &30&5 & 0&2&&     99&49(1)   	&    0&51(1)    &  7&0(1) &$\times10^{-6}$ \\
$\alpha$\textsc{-d-}Alt	&     18&5 & 0&4   & $^\mathrm{O}$&S$_2$ &     23&1 & 0&2   	&(1.17&2.87)  &45&5 & 0&3&&     99&93(1)  	&    0&065(1)   &  5&5(1) &$\times10^{-3}$ \\
$\alpha$\textsc{-d-}Ido	&     14&0 & 0&2   & $^1$&S$_3$     	 &     27&9 & 0&2   	&(1.11&4.36)  &33&6 & 0&2&&     99&47(1)  	&    0&524(3)   &  1&4(1)&$\times10^{-3}$ \\
\hline
\end{tabular}
\begin{flushleft}{ \footnotesize 
Energies are expressed in kJ/mol, angles are in rad, and probabilities in percent; $^a$ Free energy of
\invchair{}; $^b$ Next most populated conformer after \chair{} and \invchair{};
$^c$ Free energy of the next most populated conformer; $^d$ Location of the transition state ($\theta$,$\phi$);
$^e$ Free energy of the transition state ; $^f$ Population of 
\chair{} (\%); $^g$ Population of \invchair{} (\%); $^h$ Population of the next most
populated conformer. When a basin is present, which encompasses many states, all
conformers involved are listed.} 
\end{flushleft}
\caption{Free energy and population of different conformers using the 45a4 parameter set.}

\label{tab:series45a4}
\end{table*}

\begin{figure}[t]
\centering
\smallskip
\includegraphics[width=\columnwidth]{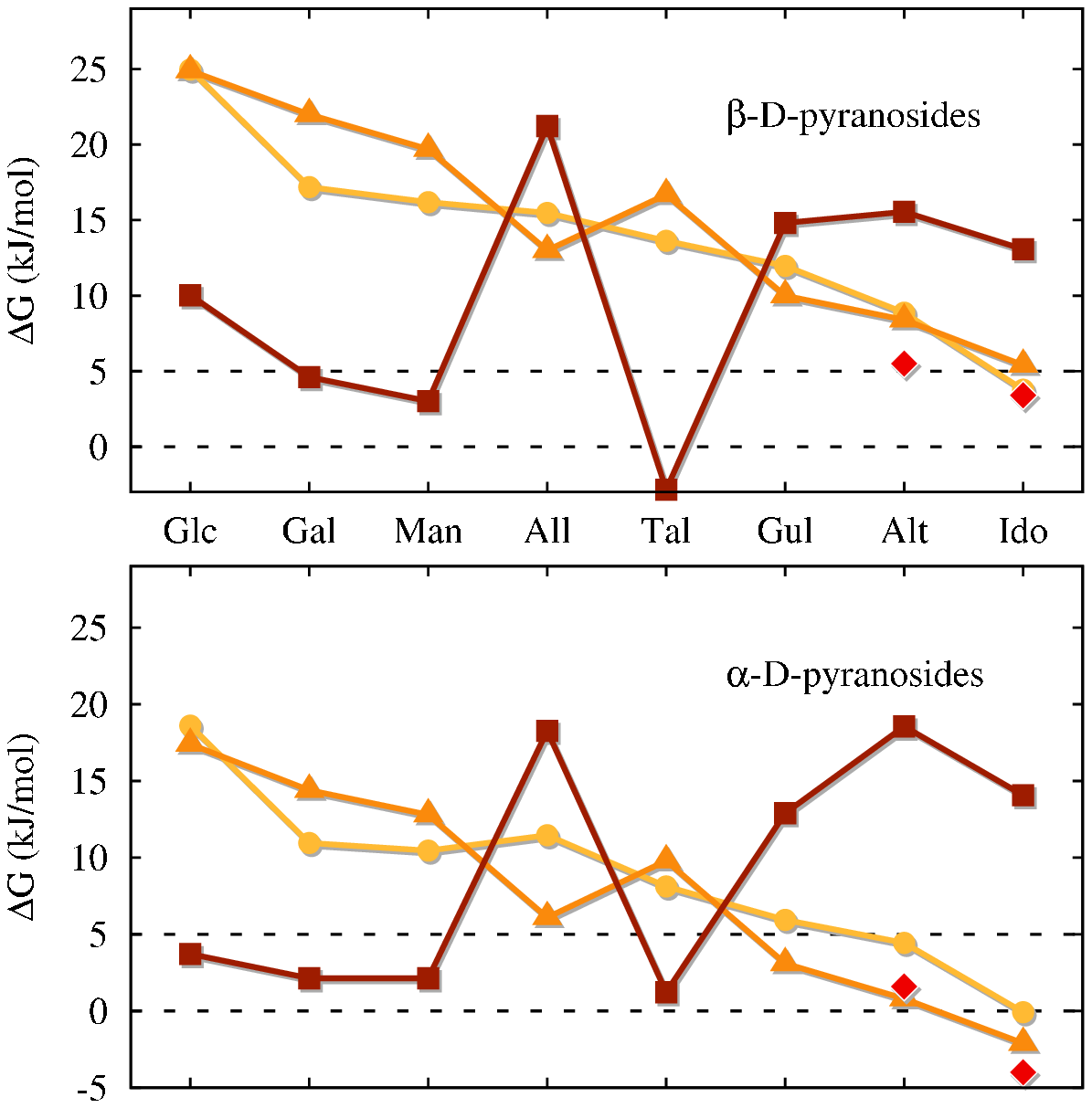}
\caption[Inverted chair free energy difference for the $\beta$ (upper panel) and
$\alpha$ (lower panel) series of aldopyranoses. The three curves refer to the
simulation results obtained using the 45a4 parameter set (squares) and the predictions of
Ref.~\onlinecite{angyal69a} (triangles) and Ref.~\onlinecite{Vijayalakshmi72} (circles). 
Diamonds corresponds to the estimates from NMR measurements (idose,
Ref.\onlinecite{Snyder86}; altrose, this work).
Lines are a guide to the eye, and error bars are always smaller than the symbols (0.5
kJ/mol).]{\label{fig:series45a4}}
\end{figure}

Simulations of the remaining 15 steroisomers of glucose have been performed using the same
protocol employed for $\beta$\textsc{-d-}Glc. We summarized the results in  Fig.~\ref{fig:series45a4} and
Tab.~\ref{tab:series45a4}. In Fig.~\ref{fig:series45a4} we report the free energy difference between the
\invchair{} and \chair{} conformers of  $\alpha$ and
$\beta$\textsc{-d-}pyranoses modeled using the 45a4 force field, along with the theoretical
estimates of of
Angyal\cite{angyal69a} and of Vijayalakshmi and coworkers\cite{Vijayalakshmi72}. Two horizontal dashed lines are also drawn at 0
and 5 kJ/mol (approximatively 2$k_\textsc{b}T$ at room temperature), highlighting the
thresholds below which the inverted chair population becomes greater than the chair one, and
below which the inverted chair population becomes noticeable, respectively. In Tab.~\ref{tab:series45a4}, free energy differences and populations
for the complete $\alpha$ and $\beta$ series are listed.
The free energy difference and population of the next
leading conformer (after \invchair) are also 
reported, as well as the location on the $(\theta,\phi)$ plane of the
first transitions next to \chair{}.  Concerning the population of the leading next conformer, the
values reported were calculated on the $\theta\in[\pi/3,2\pi/3]$ region.  This choice takes
into account all other local minima present along the equatorial line, but their free energy is always so large (as it has
been seen for $\beta$\textsc{-d-}Glc), that this  approximation does not change substantially
the population of the next leading conformer.

The differences between the theoretical estimates and the simulation results obtained using the
\textsc{gromos} 45a4 force field are striking. First of all, none of the 16 sugars investigated
presents a chair/inverted chair free energy difference in quantitative agreement with the
theory, as differences are usually larger than 5 kJ/mol, and in many cases even larger than
10 kJ/mol. More importantly, many of these values are in marked qualitative disagreement not
only with the theoretical results, but also with experimental evidence.  In fact,
$\alpha$\textsc{-d-}Glc, $\alpha$\textsc{-d-}Gal, $\alpha$\textsc{-d-}Man, $\beta$\textsc{-d-}Gal
and $\beta$\textsc{-d-}Man present an inverted chair free energy which is lower than 2 $k_\textsc{b}T$ at room temperature.  This means that a sensible population of
inverted chairs, of the order of 10\% is expected at equilibrium, in contrast with no
experimental evidence of the occurrence of this conformer.  The same behavior 
is even more pronounced in \textsc{d-}Tal, displaying an inverted chair
free energy close to zero.  On the contrary, the puckering free energy
of idose inverted chairs simulated using the 45a4 set of parameters results to be greater
than 10 kJ/mol, therefore ruling out the possibility of observing idose inverted chairs in
equilibrium simulations. 

The \textsc{gromos} 45a4 force field appears to be unable not only to compare
quantitatively with experimental and theoretical results, but --- even more importantly
--- to reproduce the qualitative behavior of any of the two series. Given the ubiquitous
presence of galactose and mannose in relevant oligo and polysaccharides of biological
origin, the inability of the force field to prevent appearance of inverted chairs at
room temperature seems to be a severe drawback, at least for out-of-equilibrium
simulations.

While the free energy of different ring conformers is certainly an
important physical quantity, one should not overlook the importance of the kinetics of the conformational
transitions. One might reason that alternate conformers might not be seen during
equilibrium simulations, if the inverse transition rate is much longer than the typical
time interval spanned by a simulation. This pragmatic approach could be hazardous, given the fast pace of increase
in simulations sizes and lengths, but nevertheless appealing. Kr\"autler and
coworkers\cite{Krautler07} reported that in 200 ns long simulation runs of
$\beta$\textsc{-d-}Glc,  $\beta$\textsc{-d-}Gal,  $\beta$\textsc{-d-}Man
and $\beta$\textsc{-d-}Tal, all sugars but glucose remained for more than 99.9\% of
the time in the chair conformation, while glucose was found in boat and twisted
conformation for the 0.7\% of the time (giving a rough estimate of the characteristic
time of escape from the chair conformer basin of 10 ns).

\begin{figure}[t]
\centering
\smallskip
\includegraphics[width=\columnwidth]{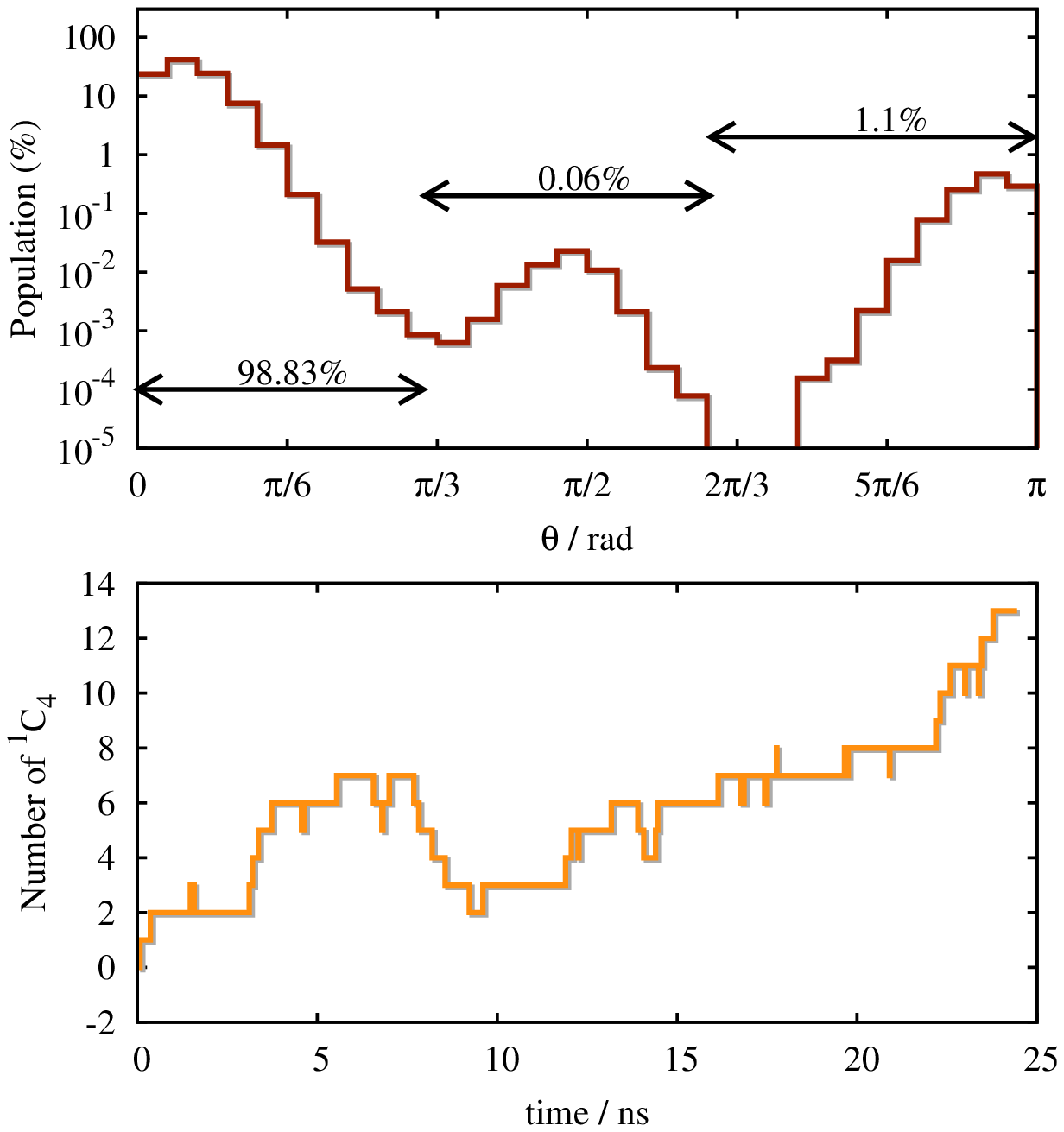}
\caption[Results of the unbiased molecular dynamics simulation of
$\beta$\textsc{-d-}Glc. Histogram of the population along the $\theta$
puckering coordinate (upper panel) and time evolution of the number of inverted chairs
 (lower panel).]{\label{fig:relaxation}}
\end{figure}

Although 200 ns is a time much longer than that of most simulations, one should keep in mind
that conformational transitions are stochastic events, and a characteristic time of 10 ns
might lead to a considerable amount of ``unwanted'' conformers in simulation runs much shorter
than 200 ns, but with more than just one sugar molecule in solution. We tested a setting which we consider
to be representative of a typical simulation of medium to large size, namely, of a 25 ns long
run at constant temperature and pressure of 512 $\beta$\textsc{-d-}Glc and
25$\times$10$^3$ water molecules in a simulation box with an edge of approximately 9.6
nm, using a cut-off of 1.3 nm for every nonbonded interaction, and an integration timestep
of 2 fs. Every other parameter and algorithm employed has been the same as in the metadynamics
run discussed so far. Indeed, we observed the appearance of both boats and inverted chairs
conformers, with a statistical frequency of 0.06\% and 1.1\%, respectively (see the upper
panel of Fig.~\ref{fig:relaxation}. These values are close to the one expected
from the free energy calculations (see Tab.~\ref{tab:series45a4}). However, a look at the time
evolution of the number of inverted chain in the simulation box (lower panel of
Fig.~\ref{fig:relaxation}) tells us that equilibrium has not been reached yet.  Still, this
result is to our opinion quite valuable, because it gives some information about the kinetics
of ring conformational transitions in $\beta$\textsc{-d-}Glc, showing that it is not unlikely to observe
inverted chair conformers in equilibrium simulations of conventional size, using the
45a4 parameter set.

We performed similar simulations of other sugars, namely, $\beta$\textsc{-d-}Gal and
$\alpha$\textsc{-d-}Glc, and in both cases we observed the appearance of inverted
chairs, although to a much smaller extent, showing that the kinetics of the chair to
inverted chair transition is much slower in these cases (consistently with the results of
Ref.~\onlinecite{Krautler07}). This fact is obviously related to the height of the barrier
that separates \chair{} conformers from other ones (see Tab.~\ref{tab:series45a4}), which appears to be lower in $\beta$\textsc{-d-}Glc than in all other
cases. 
\subsection{Force Field Re-parametrization\label{sec:reparam}}

After realizing the difficulties of the 45a4 parameter set in reproducing puckering properties
of pyranoses,  we planned to re-parametrize the force field, by finding a minimal set of changes 
that could fix at least the qualitative aspects discussed so far.
At a first glance, this task could seem a bit intimidating, because puckering variables
(and therefore the relative free energy landscapes) depend directly on all six ring atoms
and also --- indirectly, but possibly to a considerable extent --- on the ring substituents.
The number of parameters on which puckering free energy depends is, therefore, quite
high. A completely automated procedure is out of question, and one needs to adopt
an heuristic approach.

\begin{figure}[t]
\centering
\smallskip
\includegraphics[width=\columnwidth]{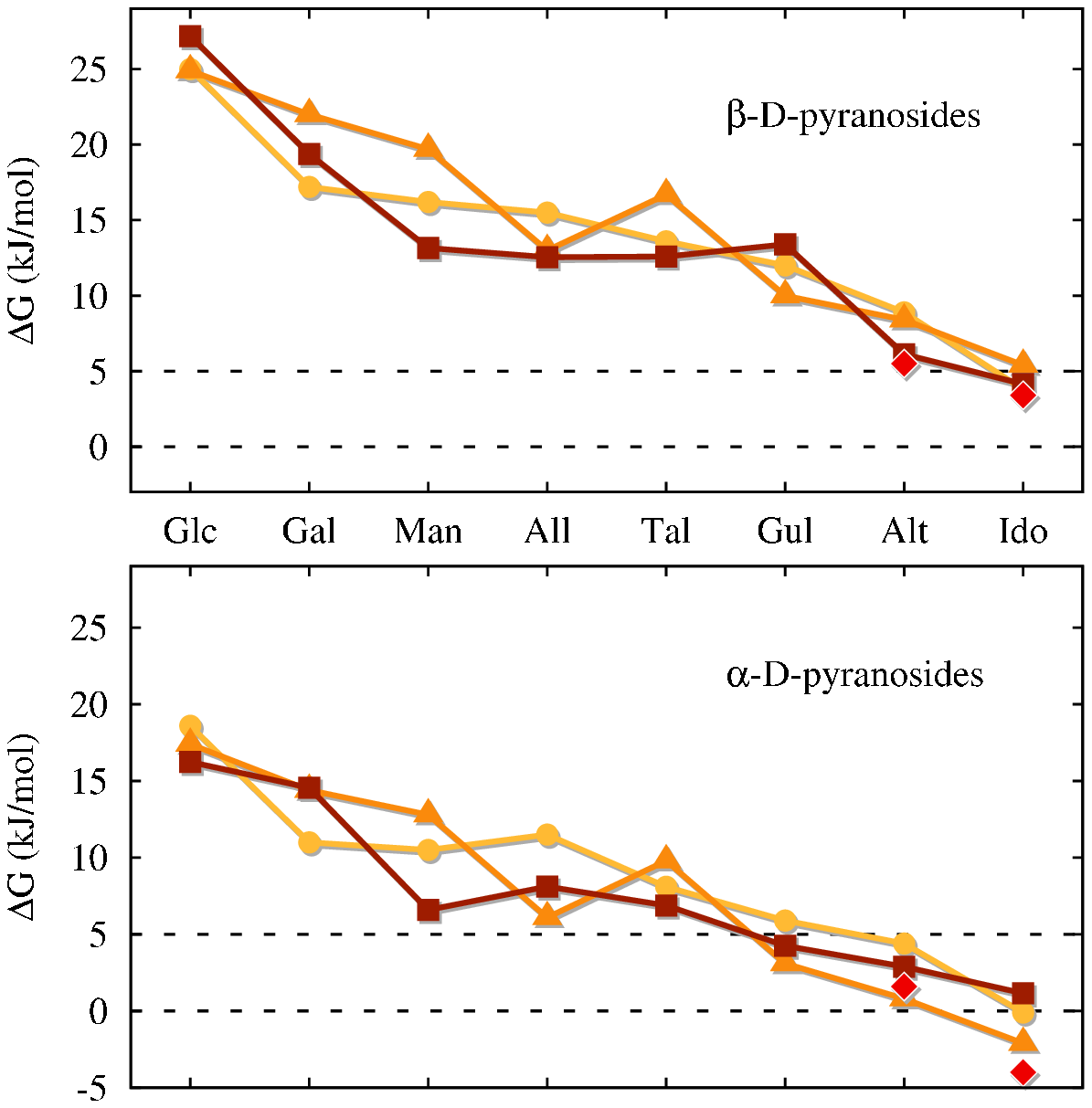}
\caption[Inverted chair free energy difference for the $\beta$ (upper panel) and
$\alpha$ (lower panel) series of aldopyranoses. The three curves refer to the
simulation results using the new parameter set (squares) and the predictions of
Ref.~\onlinecite{angyal69a} (triangles) and Ref.~\onlinecite{Vijayalakshmi72} (circles).  
Diamonds corresponds to the estimates from NMR measurements (idose,
Ref.\onlinecite{Snyder86}; altrose, this work).
Lines are a guide to the eye, and error bars are always smaller than the symbols (0.5
kJ/mol).]{\label{fig:seriesNew}}
\end{figure}

To keep the re-parametrization as general as possible, and the number of parameters to
be tuned low, we decided that our approach should adhere to the following criteria: (a) only
parameters not directly involved in inter-molecular interactions should be tuned;  (b) the
changes should not be sugar-dependent; (c) the changes have to preserve previously known or
already tuned molecular properties; (d) the inverted chair free energy of most common sugars
(glucose, galactose, mannose) should be higher than 10 kJ/mol; (e) the trend of the inverted
chair free energy as a function of the sugar type, as well as (f) the approximate offset
between the inverted chair free energies of $\alpha$ and $\beta$ anomers,
have to be reproduced. 

Within this framework, a complete, quantitative agreement for every
sugar type is most probably not feasible, especially given the constraints (a) and (b).
However, we found that quite reasonable results can be indeed achieved with minimal parameter
changes. Point (a) requires Lennard-Jones parameters and partial charges of the 45a4
parameter set to be preserved. Together with the requirement at point (d), 
this suggests that only angular or torsional interactions involving three or
more ring atoms should be the target of our optimization. This is because among the known
properties which are well reproduced by the \textsc{gromos} 45a4 force field there are the
rotameric distribution of the hydroxymethyl group and the conformation of the glycosidic
linkages (in di-saccharides). We realized quite soon that changing the stiffness of angular interactions did
not change the puckering free energy noticeably. Spieser and
coworkers\cite{Spieser99} showed that the height of the energetic barrier between \chair{} and
\invchair{} can be increased by stiffening the angular interactions. However, this
affects only the free energy of the transition states, and not the free energy
difference of the metastable conformers.  

Therefore, we concentrated on the torsional interactions, and noticed that every torsional
interaction involving three ring atoms (C1,C2,C3,C4,C5 or O5) and either C6 or any hydroxyl
oxygen (O1,O2,O3,O4) was either not present (this is the case of O4--C4--C5--O5, C3--C4--C5--C6,
O2--C2--C1--O5, C1--O5--C5--C6 and C5--O5-C1--O1) or present with a phase
term $\cos(\delta)=+1$ and multeplicity $n=2$ in the torsional potential
\begin{equation}
U(\phi)=k_\phi \left[1 + \cos(\delta)\cos(n \phi)\right],
\label{eq:torspot}
\end{equation}
associated to the dihedral angle $\phi$. As it was pointed out in Ref.~\onlinecite{hansen10a},
such a phase favors the axial conformation of the substituent, with respect to the equatorial
one, whereas a negative value of $\cos(\delta)$ would favor the equatorial conformation with
respect to the axial one.  Indeed, one of those interactions (O2--C2--C1--O5) has been
eliminated in the 45a4 version of the \textsc{gromos} force field, for the precise purpose
of stabilizing the \chair{} conformer. While this is true for glucose, it is not, \emph{e.g.},
for mannose, whose substituent in C2 is axial in the chair conformer.  Therefore, the change
in the 45a4 set that stabilized the glucose chair, acted in the opposite direction for
mannose, stabilizing the inverted chair.

\begin{table}
\preprintclearpage
\begin{tabular}{lc@{\qquad}r@{.}lcc@{\qquad}l@{\,}}
\hline
dihedral angle & & \multicolumn{2}{c}{$k_\phi$ $^\S$} &$\quad$& $\cos\delta$ & $n$ \\
\hline
\hline
C3$-$C2$-$C1$-$O1 $^\dag$       && 0&5   &&$ -1 $& 2 \\ 
C4$-$C3$-$C2$-$O2 $^\dag$       && 0&5   &&$ -1 $& 2 \\ 
C1$-$C2$-$C3$-$O3 $^\dag$       && 2&4   &&$ -1 $& 2 \\ 
C5$-$C4$-$C3$-$O3 $^\dag$       && 2&4   &&$ -1 $& 2 \\ 
C1$-$O5$-$C5$-$C6 $^\ddag$      && 0&5   &&$ -1 $& 2 \\ 
\hline
\end{tabular}
\begin{flushleft}{ \footnotesize 
$^\S$ in kJ/mol; 
$^\dag$ interaction modified with respect to the correspontent 45a4 one; 
$^\ddag$ new interaction term (not present in 45a4); The functional form for the torsional
interaction is that of Eq.\ref{eq:torspot}.  All other interactions are the same as in the 45a4 set\cite{Lins05}.
} 
\end{flushleft}
\caption{New parameters for \textsc{d-}aldopyranoses torsional interactions.} 
\label{tab:newParameters}
\preprintclearpage
\end{table}

\begin{figure*}[!htb]
\smallskip
\mbox{
\includegraphics[width=0.9\textwidth]{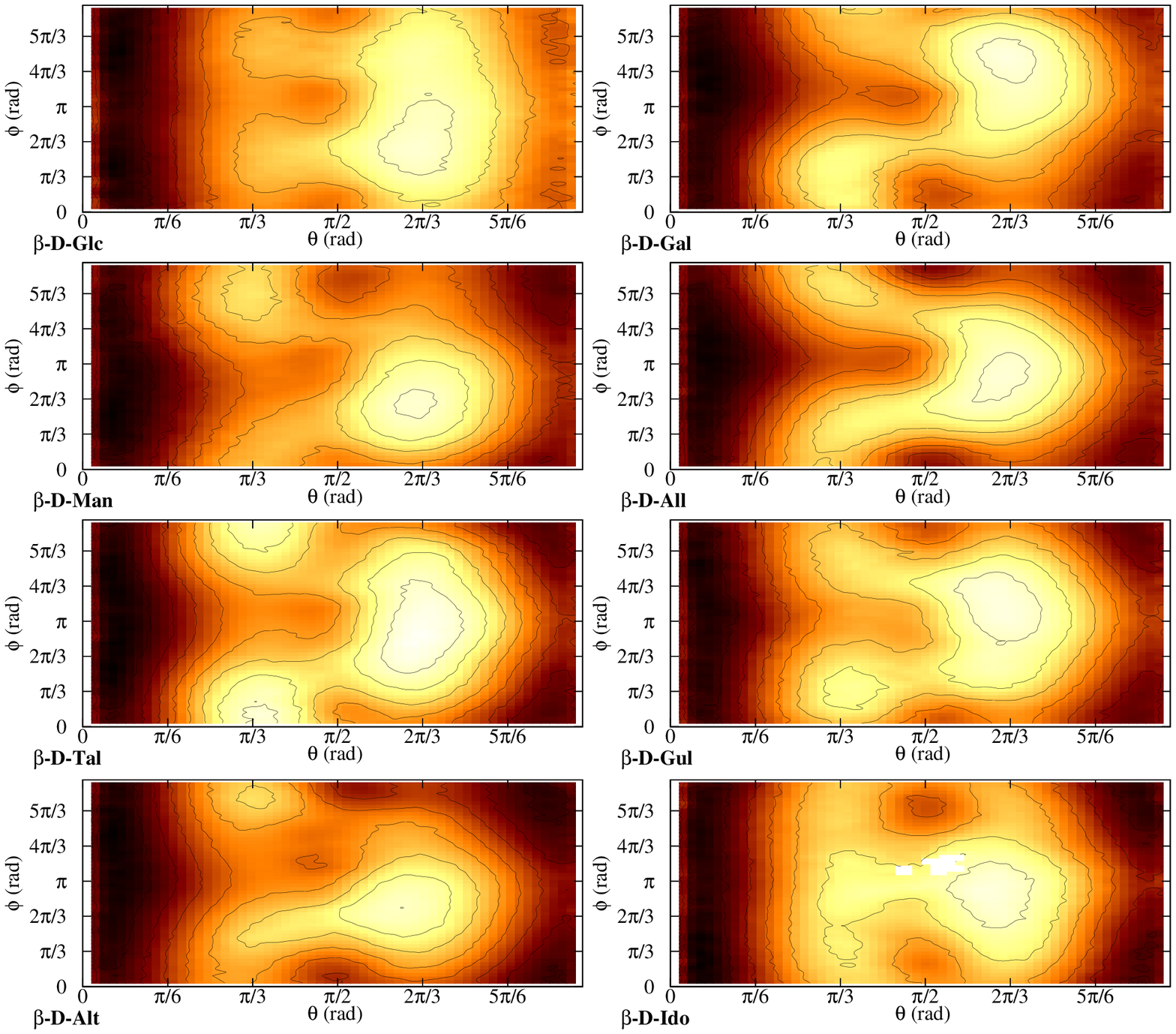}
}
\caption[
Puckering free energy landscapes (in kJ/mol) obtained by the combined metadynamics -- umbrella
sampling, for the series of $\beta$\textsc{-d-}aldopyranoses simulated using the new
force field parameters. Isolines are drawn every 10 kJ/mol, starting from the global
minimum.
]{\label{fig:NewProfilesB}}
\end{figure*}

\begin{figure*}[!htb]
\smallskip
\mbox{
\includegraphics[width=0.9\textwidth]{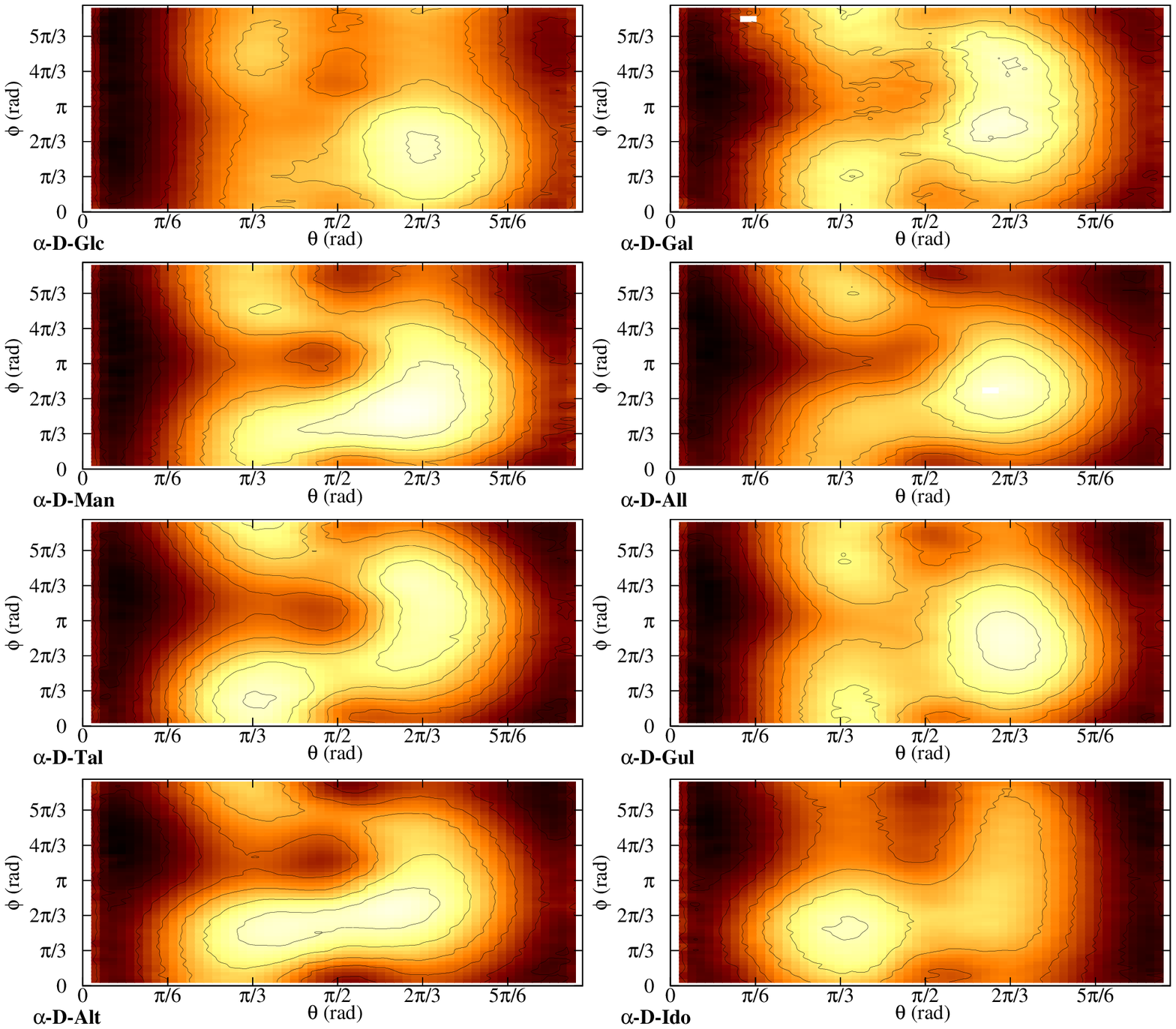}
}
\caption[
Puckering free energy landscapes (in kJ/mol) obtained by the combined metadynamics -- umbrella
sampling, for the series of $\alpha$\textsc{-d-}aldopyranoses simulated using the new
force field parameters. Isolines are drawn every 10 kJ/mol, starting from the global
minimum.
]{\label{fig:NewProfilesA}}
\end{figure*}

It is clear that any change affecting these torsional parameters will have a profound (and
sugar-specific) effect on the whole series of pyranoses. Therefore, to make grounded 
changes to the force field, one has first to understand how the pattern of axial and
equatorial substituent in the \chair{} conformer, for given torsional interactions
involving the five chiral centres, can influence the properties of the \invchair{} free energy curves of the
$\alpha$ and $\beta$ series.

Firstly, we realized that by changing the sign of $\cos(\delta)$ in the C1--C2--C3--O3 and
C5--C4--C3--O3 interactions, one can at the same time rise the \invchair{} free energy of
glucose, galactose, mannose, and talose,  and lower that of allose, gulose, altrose and
idose. Looking at the location
along the free energy series of these sugars, it appears that these interactions are leading candidates to recover the
approximate monotonous trend in the \invchair{} free energy and, consequently, to fix consequently point (e).

Another important role in determining the shape of the free energy curve is played
by a dihedral interaction which is not present in the 45a4 parameter set, namely,
the one involving the substituent at the C5 chiral center.
The chirality of C5 is the same for all 16 steroisomers of \textsc{d-}Glc, and can be
actually exploited to introduce a global shift for the \invchair{} free energies of the whole
series of 16 \textsc{d-}pyranoses.  While this gives some
freedom in our parametrization process, it should be kept in mind that, most probably, 
parameters optimized this way are not valid to model the series of \textsc{l-}pyranoses.

\begin{table*}
\begin{tabular}{l|r@{.}l@{$\,\pm\,$}r@{.}lr@{}lr@{.}l@{$\,\pm\,$}l@{.}lr@{,}lr@{.}r@{$\,\pm\,$}r@{.}lc|r@{.}lr@{.}lr@{.}l@{}l}
\hline
Isomer & \multicolumn{4}{c}{$\Delta{}G$ [\invchair{}]$^a$}&\multicolumn{2}{c}{Next}$^b$
&\multicolumn{4}{r}{$\Delta{}G$[Next]$^c$ } & \multicolumn{2}{c}{$\textrm{TS}^d$}&
\multicolumn{4}{r}{$\Delta{}G$[TS]$^e$ } &&\multicolumn{2}{c}{P[\chair{}]$^f$ }&
\multicolumn{2}{c}{P[\invchair{}]$^g$} & \multicolumn{3}{c}{P[Next]$^h$ } \\ 
 \hline 
 \hline 
 
$\beta$\textsc{-d-}Glc	&  27&1 & 0&3    & $^{3\mathrm{O}}$&B             &      30&9 & 0&2      	&(1.17&3.19)& 45&9 &0&5&&   99&99(1)  &       0&002(1)   	&  4&17(4)&$\times10^{-4}$ \\
$\beta$\textsc{-d-}Gal	&  19&3 & 0&2    & $^3$&S$_1$                     &      35&6 & 0&2      	&(1.17&3.51)& 39&7 &0&2&&   99&95(1)  &  	   0&051(1)     &  8&8(1) &$\times10^{-5}$ \\
$\beta$\textsc{-d-}Man	&  13&1 & 0&4    & $^\mathrm{O}$&S$_2$            &      32&9 & 0&4      	&(1.17&2.66)& 42&7 &0&5&&   99&58(1)  &  	   0&42(1)      &  1&41(3)&$\times10^{-4}$ \\
$\beta$\textsc{-d-}All	&  12&5 & 0&2    & $^{3\mathrm{O}}$&B             &      21&8 & 0&2      	&(1.22&3.19)& 36&8 &0&3&&   99&31(1)  &  	   0&68(1)      &  9&23(1) &$\times10^{-3}$ \\
$\beta$\textsc{-d-}Tal	&  11&9 & 0&3    & &B$_{3\mathrm{O}}$             &      43&2 & 0&1      	&(1.11&3.40)& 46&8 &0&2&&   99&53(1)  &  	   0&47(1)      &  3&0(1) &$\times10^{-6}$ \\
$\beta$\textsc{-d-}Gul	&  13&3 & 0&3    & $^\mathrm{O}$&S$_2$            &      31&5 & 0&4      	&(1.17&3.19)& 45&9 &0&5&&   99&00(2)  &  	   0&99(1)      &  3&4(1) &$\times10^{-4}$ \\
$\beta$\textsc{-d-}Alt	&  6&1 & 0&1     & $^\mathrm{O}$&S$_2$            &      24&7 & 0&1      	&(1.17&3.72)& 43&8 &0&2&&   93&88(4)  &  	   6&12(4)      &  2&4(1) &$\times10^{-3}$ \\
$\beta$\textsc{-d-}Ido	&  4&1 & 0&3     & &B$_{25}$                      &      27&4 & 0&2      	&(1.06&5.85)& 44&9 &0&2&&   88&51(9)  &  	  11&5(2)       &  8&6(1) &$\times10^{-4}$ \\
\hline\hline                                                                                                              
$\alpha$\textsc{-d-}Glc	&  16&2 & 0&2    & $^1$S$_3$ &$-^{14}$B           &      35&1 & 0&2      	&(1.22&2.66)& 42&5 &0&2&&   99&88(1)  &       0&124(1)     &  8&6(1)&$\times10^{-5}$ \\
$\alpha$\textsc{-d-}Gal	&  14&5 & 0&2    & $^1$&S$_3$                     &      44&7 & 0&2      	&(1.22&3.19)& 48&5 &0&4&&   99&58(1)  &       0&416(5)     &  4&0(1)&$\times10^{-6}$ \\
$\alpha$\textsc{-d-}Man	&  6&5 & 0&2     & $^\mathrm{O}$&S$_2$            &      22&8 & 0&1      	&(1.11&3.19)& 32&3 &0&3&&   95&72(3)  &       4&27(5)      &  6&0(1)&$\times10^{-3}$ \\
$\alpha$\textsc{-d-}All	&  8&1 & 0&2     & B$_{3\mathrm{O}} $&$- ^1$S$_3$ &      27&1 & 0&2      	&(1.17&3.40)& 38&9 &0&2&&   97&92(2)  &       2&08(3)      &  8&7(1)&$\times10^{-4}$ \\
$\alpha$\textsc{-d-}Tal	&  6&8 & 0&1     & $^\mathrm{O}$&S$_2$            &      33&7 & 0&2      	&(1.65&4.47)& 52&5 &0&2&&   91&60(6)  &       8&40(6)      &  1&7(1)&$\times10^{-4}$ \\
$\alpha$\textsc{-d-}Gul	&  4&2 & 0&2     & $^\mathrm{O}$&S$_2$            &      30&3 & 0&2      	&(1.49&3.94)& 49&2 &0&3&&   90&44(7)  &       9&6(1)       &  1&2(1)&$\times10^{-4}$ \\
$\alpha$\textsc{-d-}Alt	&  2&8 & 0&3     & &B$_{3\mathrm{O}}$             &      15&0 & 0&2      	&(1.06&3.72)& 28&9 &0&2&&   79&2(1)   &      20&7(2)       &  8&74(1)&$\times10^{-2}$ \\
$\alpha$\textsc{-d-}Ido	&  1&1 & 0&2     & $^\mathrm{O}$S$_2$ &$-^{14}$B  &      20&6 & 0&2      	&(1.11&5.53)& 31&6 &0&2&&   51&8(3)   &      48&2(2)       &  1&38(1)&$\times10^{-2}$ \\
\hline
\end{tabular}
\begin{flushleft}{ \footnotesize 
Energies are expressed in kJ/mol, angles are in rad, and probabilities in percent; $^a$ Free energy of
\invchair{}; $^b$ Next most populated conformer after \chair{} and \invchair{};
$^c$ Free energy of the next most populated conformer; $^d$ Location of the transition state ($\theta$,$\phi$); $^e$ Free energy of the transition state ; $^f$ Population of 
\chair{} (\%); $^g$ Population of \invchair{} (\%); $^h$ Population of the next most
populated conformer. When a basin is present, which encompasses many states, all
conformers involved are listed.} 
\end{flushleft}
\caption{Free energy and population of different conformers using the new
parameter set.} 
\label{tab:seriesNew}
\end{table*}

In addition to the interactions discussed so far, to reproduce
correctly the gap between the \invchair{} free energy curves of the $\alpha$ and $\beta$
series --- as it is apparent, for example, in the theoretical data presented in
Fig.~\ref{fig:series45a4} --- the torsional interaction for the C3--C2--C1--O1
present in the 45a4 set has to be modified. The chirality
of the C1 carbon atom differs only between the $\alpha$ and $\beta$ anomers, and is
certainly playing a role in modulating the height of the free energy gap  between $\alpha$ and
$\beta$ anomers. 

Eventually, corrections to the energy term associated to the dihedral
C4--C3--C2--O2 were proven to enhance the agreement to the theoretical
estimates.

In summary, the phase and amplitude of the C3--C2--C1--O1, C4--C3--C2--O2, C1--C2--C3--O3,
C5--C4--C3--O3 and C1--O5--C5--C6 torsional interaction were tuned by trial and errors in
order to obtain \invchair{} free energy curves in better agreement with experimental and
theoretical data. The set of torsional interactions and their parameters for the proposed
modification to the 45a4 parameter set are reported in Tab.~\ref{tab:newParameters}, and the full
gromacs topologies are provided at
\verb!http://www.science.unitn.it/~sega/sugars.html!
It is worth mentioning that the strength of the C1--C2--C3--O3 and C5--C4--C3--O3 torsional
interactions had to be set to a much higher value (2.4 kJ/mol) than that of the other ones
involving three ring atoms and one hydroxyl oxygen. Given the similar chemical nature of the
quadruplets of atoms involved (beside that of the anomeric oxygen), such asymmetry appears
to be peculiar. This might originate either from the fact that the O3 oxygen can be involved
in the 1,3-trans-diaxial interaction with the hydroxymethyl group (stronger than that with
any other OH groups) or from other interaction terms already present in the 45a4 set, whose
influence on the puckering free energy is not yet understood, and that might deserve a
separate investigation.

The new parametrization of the \textsc{gromos} force field for sugars was performed with the
aim not to change properties other than puckering, and dihedral interactions directly involved
in the rotameric distribution of the hydroxymethyl group were thus not changed. This choice
alone, however, is no guarantee that this quantity is not affected. We checked explicitly
that these modifications did not affect the  rotameric distribution of the hydroxymethyl
group, calculating the free energy profile of the C4--C5--C6--O6 torsional angle ($\omega$) for the
45a4 and new sets of parameters. The results obtained with the two parameter sets did not
differ more than $2\%$ between each other. The free energy surfaces have been calculated
using the same metadynamics/umbrella sampling approach employed for the calculation of
the puckering free energy, but using a Gaussian width of 0.1 rad for biasing the $\omega$ variable.

We summarized the results obtained using our modified set of parameters in
Figs.~\ref{fig:seriesNew},\ref{fig:NewProfilesB},\ref{fig:NewProfilesA} and Tab.~\ref{tab:seriesNew}. In
Fig.~\ref{fig:seriesNew} the free energy differences between inverted chair and chair
conformers are compared again with the theoretical predictions of
Ref.~\onlinecite{Vijayalakshmi72} and Ref.~\onlinecite{angyal69a}. The
improvement with respect to the results obtained with the 45a4 set
(Fig.~\ref{fig:series45a4}) is striking. Both
the $\alpha$ and $\beta$ series reproduce now the qualitative trend of the theoretical
estimates. Galactose, mannose, and talose are not anymore below the $2k_\textsc{b}T$
threshold, and the value of gulose, altrose and idose free energies diminished
considerably. Also the gap between the $\alpha$ and $\beta$ anomers is now reproduced
reasonably well, being on average $\simeq 5$kJ/mol.
Noticeably, the height of the free energy barrier of $\beta$\textsc{-d-}Glc has been
increased considerably, from $\simeq 25$kJ/mol to $\simeq 45$kJ/mol. This 
increase, which changes dramatically the kinetics of the conformational transition of
$\beta$\textsc{-d-}Glc, has no counterpart in any other $\beta-$anomer. The situation
for $\alpha-$anomers is slightly different, because the height of
the free energy barrier close to \chair{} increased markedly for  $\alpha$\textsc{-d-}Tal
and  $\alpha$\textsc{-d-}Gul and, at the same time, decreased for
$\alpha$\textsc{-d-}All and  $\alpha$\textsc{-d-}Alt. 
Unfortunately, the population of inverted chairs in $\alpha$\textsc{-d-}Ido is still
lower than that of chairs, whereas both theory and experiment show a preference
for the \invchair{} conformer. However, given the fact that these changes are
not sugar-specific, the result obtained is to our opinion still remarkable, as 
the ability to reproduce puckering properties has increased
dramatically, with respect to the 45a4 set.

\section{Conclusions\label{sec:conclusions}}

We addressed the problem of the proper modelling of pyranoses
puckering properties using the \textsc{gromos} force field. A serious
problem in such a task is the lack of experimental
information on the conformer populations of stereoisomers of glucose
other than idose. To fill this gap, we performed the first measurement
of altrose chair conformers population by assigning the complete
coupling pattern of its $^1$H NMR spectrum. The estimated populations
are in agreement within roughly 5 kJ/mol with available theoretical
analyses based on semiempirical and molecular mechanical models,
thus confirming their reliability. The theoretical estimates have
been therefore emploied as target values for the subsequent
reparametrization of the force field, for sugars other than idose
and altrose.

The realism in reproducing the population of inverted chairs of
pyranoses modeled using the 45a4 parameter set of the {\textsc
gromos} force field leaves much to be desired: galactose, mannose,
allose, and $\alpha$\textsc{-d-}Glc present a sizable population
of the inverted chair conformer, while the population of the inverted
chair conformer of idose resulted to be negligible in simulation,
contrarily to experimental evidence.

We devised a new set of parameters which has proven to be quite
successful in recovering the trend of the inverted chair free energy 
for all 16 stereoisomers under study. Our parametrization reproduces
free energy differences in accordance with experimental and theoretical
data always within 5 kJ/mol, but in most cases within 2.5 kJ/mol,
that is, $1k_\textsc{b}T$. A closer agreement with the theoretical
models is probably not needed, given the uncertainties to which
they are subjected.  Indeed, while the theoretical models do not
provide any confidence interval, an approximate picture can be
obtained by comparing them with the experimental data on
idose\cite{Snyder86} and altrose (this work), which roughly fall
within a 3 kJ/mol interval.

The improvement attained with the introduction of this new parameter set is to our opinion
substantial, and reached the goal of reproducing known puckering free energy differences
while keeping other properties, such as inter-molecular interactions and the rotameric
distribution of the hydroxymethyl group, unchanged.  These modifications to the {\sc gromos}
force field will allow to perform more realistic simulations of \textsc{d-}aldopyranoses.
They represent a certain improvement in the study of carbohydrate equilibrium properties,
but will have an even more important impact on the evaluation of out-of-equilibrium properties,
such as in the case of simulated AFM pulling experiments.

Still, much has to be done regarding the puckering properties of carbohydrates. In particular,
the role of skew conformations ---  possibly detected in NMR
experiments\cite{Snyder86,immel99a} but not significantly present in both the 45a4 and new parameter
sets --- has to be clarified. Other pyranosides and furanoses should also be investigated to further test the 
force field. 
The \textsc{gromos} force field, however, is not the only one which has been found to be problematic in reproducing
proper puckering properties, and for many force fields the investigations of ring conformer
populations are scarce, if not missing at all. Our hope is that, beside the usefluness related
to the specific case of the \textsc{gromos} force field parametrization, this work could
also serve to attract attention to the importance of the puckering problem in carbohydrate
simulations, and to stimulate further investigations.

\section*{Acknowledgements}
The authors thank E. Chiessi for enlightening discussions, and acknowledge the
use of the Wiglaf computer cluster of the Department of Physics of the University of Trento.
This work has been partially supported by a PRIN grant from the Italian Ministry of Public
Education, University and Scientific Research.
\bibliographystyle{jcp}

\makeatletter
\renewcommand{\@dotsep}{2.5}
\makeatother
\clearpage

\clearpage
\listoffigures
\listoftables
\end{document}